\documentstyle[prd,aps,epsfig]{revtex}
\newcommand{\dd}{{\rm d}}
\newcommand{\de}{{\rm e}}

\newcommand{\DD}{{\cal D}}
\newcommand{\U}{{\mathcal{U}}}

\newcommand{\Pb}{{\mathcal{P}}}
\newcommand{\Tt}{\widetilde{\mathcal{T}}}
\newcommand{\Th}{\widehat{\mathcal{T}}}
\newcommand{\Ut}{\widetilde{\mathcal{U}}}
\newcommand{\Uh}{\widehat{\mathcal{U}}}
\newcommand{\mt}{\mu}
\newcommand{\Mt}{\gamma_{_{\rm F}}}
\newcommand{\mut}{\ell}

\newcommand{\gsim}{\mathrel{%
   \rlap{\raise 0.511ex \hbox{$>$}}{\lower 0.511ex \hbox{$\sim$}}}}
\newcommand{\lsim}{\mathrel{
   \rlap{\raise 0.511ex \hbox{$<$}}{\lower 0.511ex \hbox{$\sim$}}}}
\begin{document}
\draft
\title{Localisation of massive fermions on the brane}
\author{Christophe Ringeval$^1$, Patrick Peter$^1$ and
        Jean--Philippe Uzan$^{2,1}$\vskip0.5cm}
\address{(1) Institut d'Astrophysique de Paris,
             98bis Boulevard Arago, F--75014 Paris (France)\\
         \vskip0.25cm
         (2) Laboratoire de Physique Th\'eorique, CNRS-UMR 8627,
             B\^at. 210,\\ Universit\'e Paris XI,
             F--91405 Orsay Cedex (France).}
\vskip0.25cm
\date{\today}
\maketitle
\begin{abstract}
We construct an explicit model to describe fermions confined on a four
dimensional brane embedded in a five dimensional anti-de~Sitter
spacetime. We extend previous works to accommodate massive bound
states on the brane and exhibit the transverse structure of the
fermionic fields.  We estimate analytically and calculate numerically
the fermion mass spectrum on the brane, which we show to be
discrete. The confinement life-time of the bound states is evaluated,
and it is shown that existing constraints can be made compatible with
the existence of massive fermions trapped on the brane for durations
much longer than the age of the Universe.
\end{abstract}
\pacs{Preprint number LPT-ORSAY 01-88}
\vskip2pc
\section{Introduction}\label{sec_intro}

The idea that our universe may be a hypermembrane in a five
dimensional spacetime has received some attention in the last few
years after it was realized that gravity could be localized on a
three-brane embedded in an anti-de~Sitter spacetime~\cite{RS}. Since then,
much work has been done in a cosmological context~\cite{cosmo5D} and
there is hope that a consistent (i.e., mathematically self-contained
and observationally satisfying) high dimensional model might soon be
formulated.  For instance, it has been proposed that a
model~\cite{Ekp} based on such ideas could present itself as an
alternative to the inflationary paradigm, although for the time being
the controversy as to whether or not such a model might have anything
in common with our Universe is still going on~\cite{NoEkp}.

The idea is not new however, but has evolved from the standard
Kaluza-Klein approach to that of particle localization on a higher
dimensional defect~\cite{rubakov,gibbons}. In particular, it has been
shown that massless bulk scalars and gravitons share the property to
have a zero mode localized on the brane~\cite{bajc} in the
Randall-Sundrum model. Various mechanisms~\cite{gauge0} have been
invoked according to which it would be possible to confine massless
gauge bosons on a brane, so that there is hope to achieve a reasonable
model including all the known interactions in a purely four
dimensional effective model.

A mechanism permitting localization of massless fermions on a domain
wall was described in Refs.~\cite{rubakov,jackiw}. However, although
appealing this mechanism might be, it should be emphasized that actual
fermions, as seen on an everyday basis in whatever particle physics
experiment, are massive, so that a realistic fermionic matter model on
the brane must accommodate for such a mass. The question of
localization of massive fermions on the brane thus arises naturally,
and it is the purpose of this work to provide the transverse brane and
fermionic structure that leads to this localization. Up to now,
fermions have been confined under the restricting hypothesis that the
brane self gravity was negligible~\cite{dubovski}, or that it was
embedded in a Minkowski spacetime with one~\cite{hisano} or
two~\cite{ringeval} transverse dimensions (see also~\cite{neronov} for
the localization of fermion on a string-like defect in five
dimension).

Our goal is to transpose the original work of Ref.~\cite{ringeval} to
the brane context. For that purpose, we realize the brane as a domain
wall. Such domain wall configurations in anti-de~Sitter space have
already been studied~\cite{lee,charmousis}. We will assume that five
dimensional fermions are Yukawa-coupled to the domain wall forming
Higgs field, as in the usual case of cosmic strings. In this respect,
our work somehow extends Ref.~\cite{dubovski}, where the
mass term was put by hand, and Ref.~\cite{hisano} where the gravity of
the wall was neglected.

We start, in the following section, by recalling the domain wall
configuration of a Higgs field in a five dimensional anti-de~Sitter
spacetime and discuss briefly its properties. In
section~\ref{sec_dirac}, we describe the dynamical equations of
fermions coupled to this domain wall in order to show that they obey a
Schr\"odinger-like equation with an effective potential which can trap
massive modes on the wall. The asymptotic structure, i.e., deep in the
bulk (far from the brane), is not Minkowski space, so that the
effective potential felt by the fermions possesses a local minimum at
the brane location, but no global minimum, as first pointed out in
Ref.~\cite{dubovski}. As a consequence, the bound states are
metastable and fermions can tunnel to the bulk.

We then provide an analytical approximation of the effective
potential, thanks to which we compute analytically, in
section~\ref{sec_analy}, the mass spectrum of the fermions trapped on
the brane. We obtain the mass of the heaviest fermion that can live on
the brane and estimate its tunneling rate. This result is compared to
a full numerical integration, performed in section~\ref{sec_num}. In a
last section, we investigate the parameter space and, after having
compared our results to previous ones, we conclude that there exists a
wide region in the parameter space for which the fermion masses can be
made arbitrary low, i.e., comparable to the observed small values
(with respect to the brane characteristic energy scale), while their
confinement life-time can be made much larger than the age of the
Universe. Such models can therefore be made viable as describing
realistic matter on the brane.

\section{Membrane configuration in AdS$_5$}\label{sec_wall}

We consider the action for a real scalar field $\Phi$ coupled to
gravity in a five dimensional spacetime
\begin{equation}\label{action}
S=\int\left[\frac{1}{2\kappa_{_5}^2}(R-2\Lambda)
+\frac{1}{2}g^{AB}\partial_A\Phi\partial_B\Phi
-V(\Phi)\right]\sqrt{g}\,\dd^5 x \equiv \int \sqrt{g}\,\dd^5 x [ {\cal
L}_{_{\rm grav}}+{\cal L}_\Lambda+{\cal L}_\Phi ]
\end{equation}
where $g_{AB}$ is the five dimensional metric with signature
$(+,-,-,-,-)$, $R$ its Ricci scalar, $\Lambda$ the five dimensional
cosmological constant and $\kappa_{_5}^2\equiv6\pi^2G_{_5}$, $G_{_5}$
being the five dimensional gravity constant. Capital Latin indices
$A,B\ldots$ run from 0 to 4. The potential of the scalar field $\Phi$
is chosen to allow for topological membrane (domain wall like)
configurations,
\begin{equation}\label{V}
V(\Phi)=\frac{\lambda}{8}\left(\Phi^2-\eta^2\right)^2,
\end{equation}
where $\lambda$ is a coupling constant and $\eta=\langle
|\Phi|\rangle$ is the magnitude of the scalar field vacuum expectation
values (VEV)\footnote{Note, that, because of the unusual number of
spacetime dimensions, the fields have dimensions given by $[R]=M^2$,
$[\Phi]=M^{3/2}$, $[\Lambda]=M^2$, $[\lambda]=M^{-1}$,
$[\eta]=M^{3/2}$ and $[\kappa_5]=M^{-3/2}$ ($M$ being a unit of
mass).}.

Motivated by the brane picture, we choose the metric of the bulk
spacetime to be of the warped static form
\begin{equation}\label{metric}
\dd s^2=g_{AB} \dd x^A \dd x^B = -\dd y^2 +
\hbox{e}^{-2\sigma(y)}\eta_{\mu\nu} \dd x^\mu\dd x^\nu = -\dd y^2 +
g_{\mu\nu} \dd x^\mu\dd x^\nu,
\end{equation}
where $\eta_{\mu\nu}$ is the four dimensional Minkowski metric of
signature ($+,-,-,-$), and $y$ the coordinate along the
extra-dimension. Greek indices $\mu,\nu\ldots$ run from 0 to 3.

With this metric ansatz, the Einstein tensor components reduce to
\begin{equation} G_{\mu\nu} = -g_{\mu\nu} (6\sigma'^2 - 3\sigma''), \
\ \ G_{yy} = -6\sigma'^2,\end{equation} where a prime denotes
differentiation with respect to $y$. The non-vanishing components of
the matter stress-energy tensor
\begin{equation}T_{AB} \equiv 2{\delta {\cal L}_\Phi \over \delta
g^{AB}} - g_{AB} {\cal L}_\Phi
\end{equation}
are given by
\begin{equation} T_{\mu\nu} = {1\over 2} g_{\mu\nu} (\Phi'^2 + 2 V), \
\ \ \ T_{yy} = {1\over 2} (\Phi'^2 - 2 V).\label{tmunu}\end{equation}
It follows that the five dimensional Einstein equations
\begin{equation} G_{AB} + \Lambda g_{AB} = \kappa_5^2 T_{AB}
\end{equation}
can be cast in the form
\begin{eqnarray}
\frac{3}{\kappa_{_5}^2}\sigma'' & = &
\Phi^{\prime2}\label{einstein1}\\ 6\sigma^{\prime 2}& =
&\frac{\kappa_{_5}^2}{2}\left(\Phi^{\prime2}-2V\right) - \Lambda ,
\label{einstein2}
\end{eqnarray}
while the Klein--Gordon equation takes the form
\begin{equation}\label{kg}
\Phi''-4\sigma'\Phi'=\frac{\dd V}{\dd\Phi} .
\end{equation}
Eqs.~(\ref{einstein1}-\ref{kg}) is a set of three differential
equations for two independent variables ($\Phi$ and $\sigma$). Indeed,
as can easily be checked, the Klein-Gordon equation stems from the
Einstein equations provided $\Phi'\neq 0$. To study the domain wall
configuration, we choose to solve the first Einstein equation
(\ref{einstein1}) together with the Klein-Gordon equation (\ref{kg}).

This system of equations must be supplemented with boundary
conditions. By definition of the topological defect like
configuration, we require that the Higgs field vanishes on the
membrane itself, i.e., $\Phi = 0$ for $y=0$, while it recovers its VEV
in the bulk, so that $\lim_{y\to\pm\infty} \Phi = \pm \eta$. Note that
the sign choice made here is arbitrary and corresponds to the
so-called kink solution~; the opposite choice
(i.e. $\lim_{y\to\pm\infty} \Phi = \mp \eta$) would lead to an
anti-kink whose physical properties, as far as we are concerned, are
exactly equivalent. As for the metric function $\sigma$, it stems from
the requirement that one wants to recover anti-de~Sitter
asymptotically, so that one demands that $\sigma'$ tends to a constant
for $y\to\pm\infty$. This constant can be determined using
Eq.~(\ref{einstein2}), so that $\lim_{y\to\pm\infty} \sigma' = \pm
\sqrt{-\Lambda/6}$. Note that as $y$ changes sign at the brane
location, there is no choice for the sign of the function $\sigma$ in
this case. Note also that, as is well known, the static hypothesis
implies that the bulk cosmological constant $\Lambda$ must be
negative, and therefore the five dimensional spacetime to be
anti-de~Sitter.

With the convenient dimensionless rescaled variables
\begin{equation} \varrho\equiv  y\sqrt{|\Lambda|},\ \ \ \ H\equiv
\frac{\Phi}{\eta},\ \ \ \ \ S\equiv\frac{\dd\sigma}{\dd \varrho},
\end{equation}
the dynamical equations read
\begin{equation} \dot S = \frac{\alpha}{3} \dot
H^2,\label{eq1}\end{equation}
\begin{equation} \ddot H -4S\dot H = 4\beta H (H^2 -1),\label{eq2}
\end{equation}
where a dot refers to a derivative with respect to $\varrho$ and the
two dimensionless (positive) parameters $\alpha$ and $\beta$ are
defined by
\begin{equation} \alpha \equiv \kappa_{_5}^2 \eta^2,\ \ \ 
\beta \equiv \frac{\lambda \eta^2}{8 |\Lambda|}.
\end{equation}

These parameters are not independent since, for an arbitrary value of
$\beta$ say, there is only one value of $\alpha$ for which the
boundary condition $S (0)= 0$, or equivalently $\lim_{\varrho\to
-\infty} S(\varrho) = -1/\sqrt{6}$, is satisfied. This stems from the
fact that Eq.~(\ref{eq1}) is a first order equation in $S$, so that
only one boundary condition is freely adjustable, and we choose it to
be at $\varrho\to+\infty$.  Once this choice is made, the value of $S$
on the brane is completely determined, and unless the parameters are
given the correct values, it does not vanish. As the solution must be
symmetric with respect to the extra dimension coordinate $y$, one must
tune the parameters in order to have a meaningfull solution (i.e. for
which the metric and its first derivative are continuous at
$\varrho=0$). This is reminiscent of the relation that should hold
between the brane and bulk cosmological constants~\cite{charmousis}.

\begin{figure}[t]
\begin{center}
\epsfig{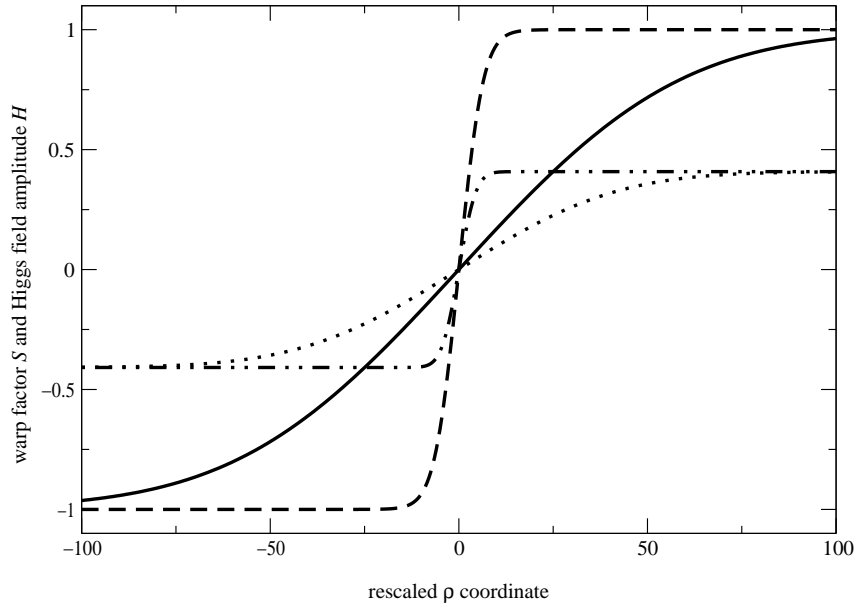}
\vskip6mm
\caption{The rescaled Higgs field amplitude $H$ (full line for $\beta
= 0.01$ and dashed line for $\beta =0.1$) and warp factor derivative
$S$ (dotted line for $\beta=0.01$ and dot-dashed line for $\beta=0.1$)
as functions of the rescaled extra dimension coordinate $\varrho$.}
\label{fig3}
\end{center}
\end{figure}

Eqs.~(\ref{eq1}~-\ref{eq2}) have been solved numerically with the
relevant boundary conditions. The field profiles are depicted on
Fig.~\ref{fig3} for two arbitrary values of the parameter $\beta$.

\begin{figure}[t]
\begin{center}
\vskip9mm
\epsfig{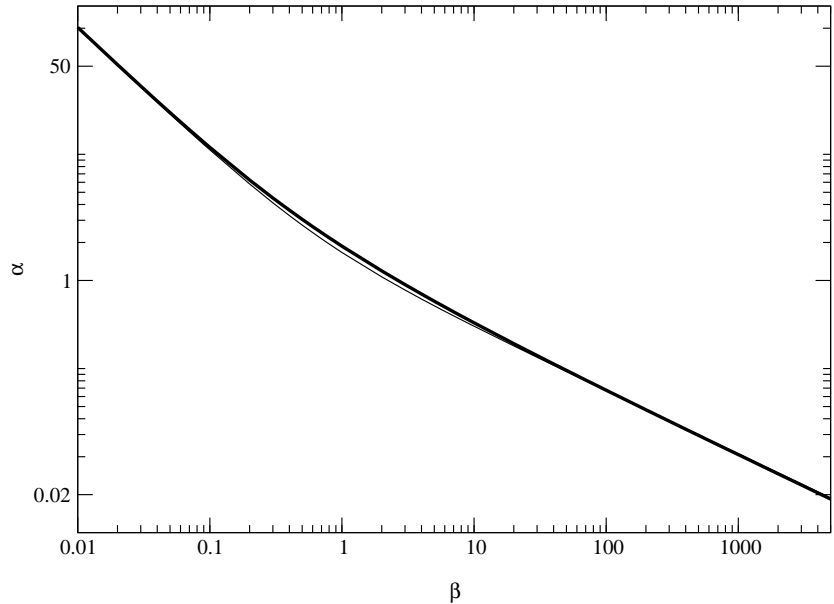}
\vskip6mm
\caption{Relation between the dimensionless constants $\alpha$ and
$\beta$. The thick curve represent the result of the numerical
integration, while the (hardly distinguishable) thin curve is the best
analytical fit.}
\label{RPU2}
\end{center}
\end{figure}

The relation between the parameters $\alpha$ and $\beta$ such that the
metric is regular at the brane location is depicted on
Fig.~\ref{RPU2}.  As can be seen on the figure, it consists
essentially in two power laws. For small values of $\beta$, one finds
roughly $\alpha\sim 1/\beta$, which becomes exact in the limit
$\beta\to 0$, while for large values of $\beta$, one gets $\alpha\sim
{4\over 3} \beta^{-1/2}$, which is, again, exact in the limit
$\beta\to\infty$.  We were able to find the best fit
\begin{equation}
\alpha^2 = {1\over \beta} \left[ {1\over \beta} + \left( {4\over 3}
\right)^2 \right],
\label{relab}
\end{equation}
which, as can be seen on the figure, is almost exact everywhere. This
translates into the relation 
\begin{equation} |\Lambda | = {1\over 9}\lambda\eta \left(
\sqrt{1+\left( {9\kappa_5^2\eta^2\over
8}\right)^2}-1\right)\label{rellambda}\end{equation} between the 5
dimensional cosmological constant and the microscopic parameters,
which corresponds to the usual relations between brane and bulk
cosmological constants.

\section{Localization of fermions on the wall}\label{sec_dirac}

This section is devoted to the description of the Dirac equation in
five dimensions with the domain wall configuration obtained in the
previous section.

The minimal representation of spinors in five dimensions can be chosen
to be four dimensional~\cite{polchinsky}. The five dimensional
Clifford algebra can then be constructed from the usual four
dimensional one by adding the $\gamma_5$ matrix to close the
algebra. If $\gamma^\mu$ design the usual four dimensional Dirac
matrices in Minkowski space in the chiral representation, the Dirac
matrices in five dimensional Minkowski space, $\Gamma^A$, are
\begin{eqnarray}\label{def_dirac}
\Gamma^\mu =  \gamma^\mu, \qquad \Gamma^4 = -i \gamma_5
\end{eqnarray}
and they satisfied the usual Clifford algebra
\begin{equation}\label{clifford}
\{\Gamma^A,\Gamma^B\}=2 \eta^{A B}
\end{equation}
where $\eta^{A B}$ stands for the five dimensional Minkowski
metric. Since in this representation the Dirac matrices satisfy
\begin{equation}\label{Gamma_prop}
\Gamma^5=i\Gamma^0\dots \Gamma^4=\mathrm{Id}
\end{equation}
the five dimensional spinors have neither Weyl nor Majorana
represention. It follows that the Dirac Lagrangian in five dimension
for fermions coupled to the Higgs domain wall is necessary of the
form
\begin{equation}\label{lagrangien}
{\mathcal{L}}_\psi= \sqrt{g} \left( i \overline\Psi \Gamma^A \DD_A \Psi
- g_{_{\rm F}} {\overline\Psi} \Phi \Psi \right) = 
\sqrt{g}\, {\overline\Psi}\left[i
\hbox{e}^{\sigma(y)} \Gamma^\mu \DD_\mu + i \Gamma^4
\partial_y  -g_{_{\rm F}}\Phi\right]\Psi
\end{equation}
where the Lorentz covariant derivative with  spin connection
is~\cite{birrel}
\begin{equation}\label{derivative}
\DD_\mu\equiv\partial_\mu-\frac{1}{2} \sigma'(y) \hbox{e}^{-\sigma(y)}
\Gamma_\mu \Gamma^4.
\end{equation}
We emphasize that the sign of the coupling $g_{_{\rm F}}$ of the
spinor $\Psi$ to the Higgs field is arbitrary and represents a
coupling either to kink or to anti-kink domain wall. For definiteness,
we shall consider in what follows only the case of a kink coupling,
and thus assume without lack of generality that $g_{_{\rm
F}}>0$\footnote{Note also that the dimensions are given by $[\Psi ]=M^2$ and
$[g_{_{\rm F}} ]= M^{-1/2}$.}.

The variation of the Lagrangian (\ref{lagrangien}) leads to the
equation of motion of the spinor field, namely the Dirac equation in
five dimensional anti-de~Sitter space for a fermionic field coupled to
a Higgs field,
\begin{equation}\label{fermmvt}
\left[i \Gamma^4\left(\partial_y- 2\sigma'(y)\right) +
i \hbox{e}^{\sigma(y)} \Gamma^\mu\partial_\mu
 - g_{_{\rm F}} \Phi \right] \Psi=0.
\end{equation}
This equation involves the matrix $\gamma_5$ (through $\Gamma_4$)
and it is thus convenient to split the four dimensional right-
and left-handed components of the five dimensional spinor and to
separate the variables as
\begin{equation}\label{sepansatz}
\Psi(x^\mu,y)= \left[ \frac{1+\gamma_5}{2}\U_{\rm
R}(y)+\frac{1-\gamma_5}{2}\U_{\rm L}(y)\right] \psi(x^\mu),
\end{equation}
where $\psi(x^\mu)$ is a four dimensional Dirac spinor, while $\U_{\rm
R}(y)$ and $\U_{\rm L}(y)$ are yet-underdetermined functions of $y$.
In what follows, we want the five dimensional Dirac equation to yield
an effective four dimensional massive Dirac equation, with an
effective mass $m$ (energy eigenvalue of the bound state).
Such a requirement implies that
\begin{equation}
i \gamma^\mu \partial_\mu \psi =m \psi,
\end{equation}
or, equivalently, in terms of the right- and left- handed
components
\begin{equation}\label{diracmass}
i\gamma^\mu \partial_\mu \psi_{\rm R} =  m \psi_{\rm L},\qquad
i\gamma^\mu \partial_\mu \psi_{\rm L} = m \psi_{\rm R},
\end{equation}
where the right- and left-handed components of the four
dimensional spinor are defined as
\begin{equation}\label{leftright4d}
\psi_{\rm R}\equiv\frac{1+\gamma_5}{2}\psi\qquad \textrm{and}\qquad
\psi_{\rm L}\equiv\frac{1-\gamma_5}{2}\psi.
\end{equation}
Contrary to the case studied in Ref.~\cite{dubovski}, the mass $m$ is
not an arbitrary parameter and will be determined later.

Choosing $\psi_{\rm R}$ and $\psi_{\rm L}$ as the independent
variables instead of $\Psi$ and $\overline\Psi$, and inserting
equation (\ref{diracmass}) into the equation of motion (\ref{fermmvt})
while using the splitting ansatz (\ref{sepansatz}) yields the
differential system for the two functions $\U_{{\rm R}/{\rm L}}(y)$,
\begin{eqnarray}\label{systurul}
\left[\partial_y-2 \sigma'(y)-g_{_{\rm F}} \Phi\right]\U_{\rm R}(y)&=&
-m\de^{\sigma(y)}\U_{\rm L}(y), \label{systurul1} \\
\left[\partial_y-2 \sigma'(y)+g_{_{\rm F}} \Phi\right]\U_{\rm L}(y)&=&
m\de^{\sigma(y)}\U_{\rm R}(y). \label{systurul2}
\end{eqnarray}
To simplify the notations, it is convenient to introduce the
dimensionless rescaled bulk components of the fermions
\begin{equation}\label{defut}
\Ut(\varrho)\equiv\de^{-\frac{3}{2}\sigma(\varrho)}
\frac{\U(\varrho)}{|\Lambda|^{1/4}},
\end{equation}
in terms of which the system (\ref{systurul1}-\ref{systurul2})
takes the form
\begin{eqnarray}
\left(\partial_\varrho - \left[\Mt H + \frac{1}{2}
S \right]\right)\, \Ut_{\rm R} &=& - \mt\, \de^{\sigma}\Ut_{\rm L},
\label{syst1}\\
\left(\partial_\varrho + \left[\Mt H - \frac{1}{2}
S\right]\right)\, \Ut_{\rm L} &=& \mt\, \de^{\sigma}\Ut_{\rm R},
\label{syst2}
\end{eqnarray}
with the dimensionless rescaled mass and coupling constant
\begin{eqnarray}\label{reduce_para}
\mt\equiv\frac{m}{\sqrt{|\Lambda|}} \quad \textrm{and} \quad
\Mt\equiv\frac{g_{_{\rm F}} \eta}{\sqrt{|\Lambda|}}.
\end{eqnarray}
Let us first concentrate on the special case $\mt=0$. The system
(\ref{syst1}-\ref{syst2}) then consists in two decoupled differential
equations and the zero mode states~\cite{randjbar} are
recovered. Asymptotically, these functions behave as
\begin{eqnarray}
\Ut_{\rm R}(\varrho\rightarrow\pm\infty) & \sim & \de^{\left(\Mt +
\frac{1}{2\sqrt{6}}\right)|\varrho|},\\ \Ut_{\rm
L}(\varrho\rightarrow\pm\infty) & \sim & \de^{-\left(\Mt -
\frac{1}{2\sqrt{6}} \right)|\varrho|}.
\end{eqnarray}
Thus, only the left-handed solution $\Ut_{\rm L}$ may remain
bounded~\cite{bajc,jackiw}, and yet provided
\begin{equation}\label{cond_bounded}
\Mt > \frac{1}{2\sqrt{6}}.
\end{equation}
Indeed, the right-handed zero modes could have been obtained by
considering the coupling of fermions to the anti-kink Higgs
profile\footnote{The coupling with and anti-kink for which $g_{_{\rm
F}}<0$ would have yield right-handed solution with the constraint $\Mt
< -\frac{1}{2\sqrt{6}}.$}. We thus recover the well-known fact that
massless fermions {\it must} be single-handed in a brane model,
contrary to the ordinary four dimensional field theory in which they
simply {\it can}.

Let us now focus on the more interesting massive case for which
$\mt\not=0$. Then the system (\ref{syst1}-\ref{syst2}) can be
decoupled by eliminating $\Ut_{\rm R}$ say. For that purpose, we
differentiate (\ref{syst2}) with respect to $\varrho$ and express
$\partial_\varrho\Ut_{\rm R}$ using equation (\ref{syst1}) and
$\Ut_{\rm R}$ using equation (\ref{syst2}) again to get
\begin{eqnarray}\label{ordertwoul}
&&\left[\partial^2_\varrho - 2 S \,\partial_\varrho + \left(\mt^2
\de^{2 \sigma} +\frac{3}{4} S^2 - \Mt^2 H^2 - \Mt S H
- \frac{1}{2} \partial_\varrho S +\Mt  \partial_\varrho H
\right)\right] \Ut_{\rm L}= 0,\\ 
\label{orderzerour}
&& \Ut_{\rm R} = \frac{\de^{-\sigma}}{\mt}\left[\partial_\varrho +
\left(\Mt H - \frac{1}{2} S \right)\right]\Ut_{\rm L}.
\end{eqnarray}
This system is strictly equivalent to the initial system
(\ref{syst1}-\ref{syst2}) since differentiating
Eq.~(\ref{orderzerour}) and then using Eq.~(\ref{ordertwoul}) gives
back Eq.~(\ref{syst1}). It is thus important to keep both
equations. Note that the integration of the first equation
(\ref{ordertwoul}) will require two initial conditions but that
$\Ut_{\rm R}$ will then be completely determined and thus requires no
extra constant of integration. As a consequence, it is sufficient to
solve the second order equation (\ref{ordertwoul}) for $\Ut_{\rm L}$
in order to fully determine the left- and right-handed bulk fermion
profiles.

Eq.~(\ref{ordertwoul}) can be recast into a Schr\"odinger-like second
order differential equation
\begin{equation}
\label{schrodlike}
\partial^2_\varrho \,\Uh_{\rm L} + \omega^2(\varrho)\, \Uh_{\rm L} = 0,
\end{equation}
where the function $\omega$ is defined by
\begin{equation}\label{omega2}
\omega^2(\varrho) \equiv \mt^2 \de^{2 \sigma(\varrho)} +
\partial_\varrho\left(\Mt H + \frac{1}{2} S \right) - 
\left(\Mt H+\frac{1}{2} S\right)^2
\end{equation}
with the new function
\begin{equation}\label{defuh}
\Uh(\varrho)\equiv\de^{-\sigma(\varrho)} \Ut(\varrho).
\end{equation}
Our aim will now be to find the zero modes of this new equation; as
previously discussed, they will be equivalent to the massive bound
states we are looking for on the brane.

\begin{figure}[t]
\begin{center}
\epsfig{file=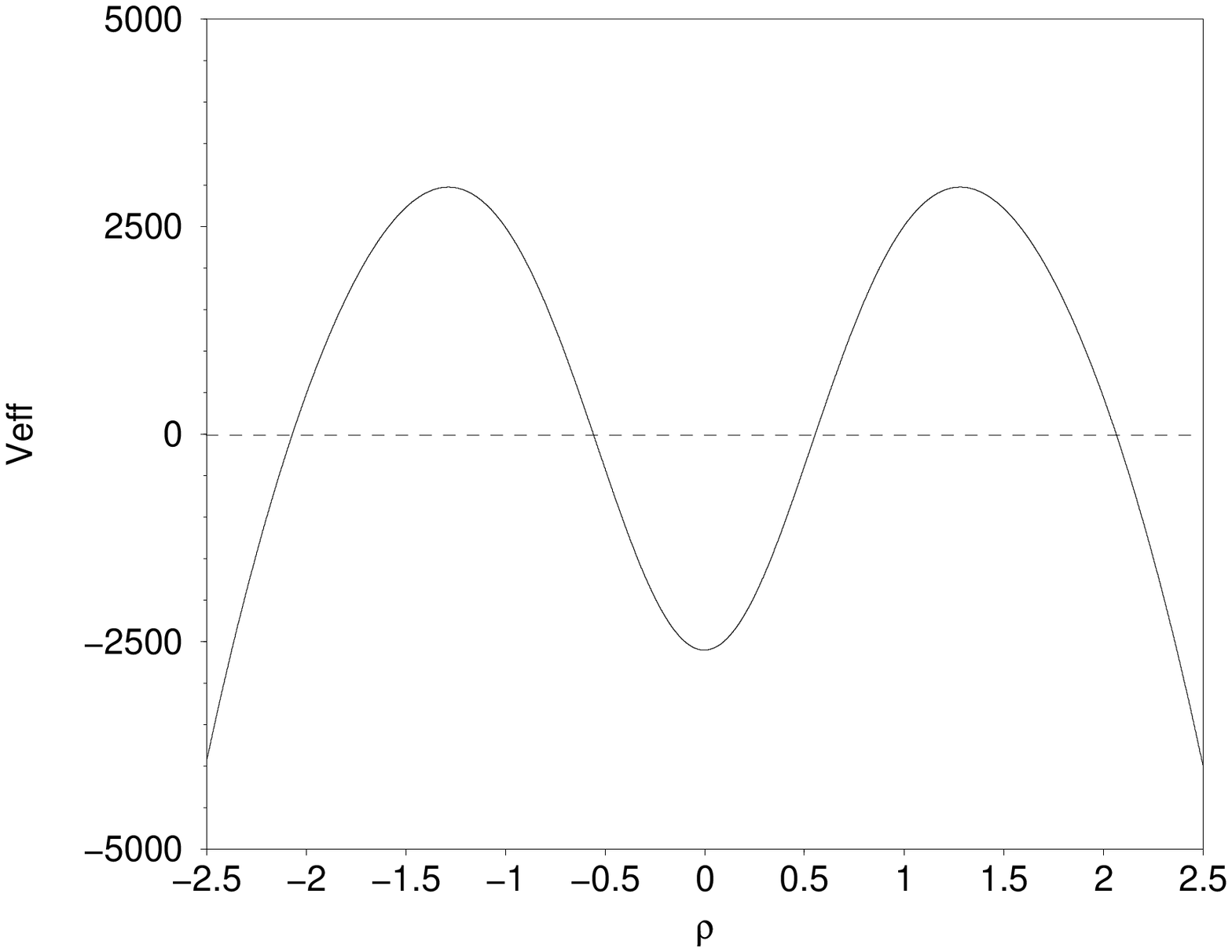,width=8cm}
\caption{The effective potential $V_{_{\rm eff}}=-\omega^2$ felt by
$\Uh_{\rm L}$. We chose the parameters $\Mt= 100$ and $\mt = 0.5 \Mt$
(an illustrative value, not necessarily leading to a bound state) in
the Higgs and gravity background similar to those of Fig.~\ref{fig3},
and we have assumed, for definiteness, the values $\alpha =1.63$,
$\beta=1.23$. The fermions are trapped on the brane where $V_{_{\rm
eff}}$ is negative and have a non-zero probability of tunneling into
the bulk due to combined effects of Higgs and gravity which produce a
finite potential barrier.}\label{figveff}
\end{center}
\end{figure}

In order for the fermions to be confined on the brane, the minimum of
$\omega^2$ needs to be negative to imply an exponential decrease of
$\Uh_{\rm L}$ in the bulk. This is essentially equivalent to the
condition (\ref{cond_bounded}) that was obtained for the case of zero
modes. We shall assume henceforth that this condition also holds for
massive modes, i.e., that the value of $\Mt$ necessary to bind massive
fermions on the brane is at least that to bind massless ones. Indeed,
Eq.~(\ref{omega2}) shows that the minimum of $\omega^2$ can only be
negative for large values of the parameter $\Mt$. However,
since the first term of (\ref{omega2}) increases exponentially at
large distance from the brane, then, if $\mt\not=0$, $\omega^2$ will
necessarily become positive.  This will yield asymptotic radiative
behaviors of the spinor bulk components. Physically, it can be
interpreted as a tunneling of the fermions from the brane to the
bulk~\cite{dubovski}. On the other hand, on the brane, the Higgs field
$H$ and the derivative of the warp factor $S$ vanish, so that
$\omega^2(0)$ is positive. As a result, the fermions can freely
propagate in a tiny region around the brane, but certainly only for
particular values of $\omega^2$ (and thus of $\mt$) satisfying the
boundary conditions with the surrounding exponential decreasing
regions. The effective potential $V_{_{\rm eff}}=-\omega^2$, depicted
in Fig.~\ref{figveff}, exhibits a local minimum on the brane and
minima at infinity.  The modes trapped on the brane are thus expected
to have discrete masses $\mt$ on the brane and non-zero probability of
tunneling into the bulk.

\section{Analytic estimate of the mass spectrum and of
the tunneling rate}\label{sec_analy}

Since the Higgs and warp factor profiles are not known analytically,
it is a priori impossible to solve Eq.~(\ref{schrodlike})
analytically. Nevertheless, since the fermions are expected to be
trapped in a neighborhood of the brane, we can look for series
solutions in $\varrho$.

In \S~\ref{subsec_approx}, we give an approximation of the effective
potential $V_{_{\rm eff}}$ which will then be used to determine the
bound states and the mass spectrum in \S~\ref{subsec_bound}. We end
this section by determining in \S~\ref{subsec_tunel} the tunneling
rate in this approximation. The validity of this approach is difficult
to assess and it will be justified {\sl a posteriori} on the ground of
a full numerical integration of the system in the following section.
In the whole section, it is assumed that $\mt>0$.

\subsection{Approximation of the effective potential}\label{subsec_approx}

In a neighborhood of the brane we can expand
the Higgs and warp factor profiles as
\begin{eqnarray}
H(\varrho) & = & H_1\varrho + H_3\varrho^3 +
{\mathcal{O}}(\varrho^4),\label{fieldexpand1}\\
S(\varrho) & = & S_1\varrho + S_3 \varrho^3 + {\mathcal{O}}(\varrho^4),
\label{fieldexpand2}
\end{eqnarray}
where both the constant and quadratic terms vanish for symmetry
reasons. To simplify the analysis, we shall make use of the equations
of motion in the form
\begin{equation} \dot S  = {2\over 3} F(S,H),\label{dotS}
\end{equation}
and
\begin{equation} \dot H  = \pm \sqrt{{2\over \alpha} F(S,H)},\label{dotH}
\end{equation}
where the function $F(S,H)$ is defined as
\begin{equation}F(S,H) \equiv 6 S^2 +\alpha\beta (H^2-1)^2 -1, \label{F}
\end{equation}
(recall that $\Lambda < 0$).

Plugging the expansions (\ref{fieldexpand1}) and (\ref{fieldexpand2})
into Eqs.~(\ref{dotS}) and (\ref{dotH}), it follows that the function
$F(S,H)$ can be expanded up to third order as
\begin{equation}\label{Fexpand}
F(S,H) \sim \alpha \beta - 1 + \left(6 S_1^2 - 2 \alpha \beta H_1^2
\right) \varrho^2 + {\mathcal{O}}(\varrho^4).
\end{equation}
Inserting this expression back into the equations of motion yields the
three coefficients
\begin{eqnarray}
S_1 & = &\frac{\alpha}{3}H_1^2,\label{coeffs1}\\
S_3 & = & \frac{4}{9} \alpha H_1^2
\left[\frac{\alpha}{3} H_1^2 - \beta \right],\label{coeffs2}\\
H_3 & = & \frac{2}{3} H_1 \left[\frac{\alpha}{3} H_1^2 -
\beta \right]\label{coeffs3}
\end{eqnarray}
in terms of the coefficient $H_1$
\begin{equation}\label{def_H1}
H_1\equiv\left.\partial_\varrho H\right\vert_{\varrho=0}=
\sqrt{\frac{2}{\alpha}(\alpha\beta-1)}.
\end{equation}
Then, the frequency $\omega^2$ can be expanded as an harmonic
oscillator potential
\begin{equation}\label{omega2near}
\omega^2(\varrho)=\omega_{\rm b}^2(\varrho)+{\mathcal{O}}(\varrho^4)\qquad
\hbox{with}\qquad
\omega^2_{\rm b}(\varrho)\equiv\omega_0^2 - \Omega \,\varrho^2
\end{equation}
where $\omega_0^2$ and $\Omega$ are given by
\begin{eqnarray}
\omega_0^2 &\equiv& \mt^2 + \Mt H_1\left(1 + \frac{\alpha}{6 \Mt} H_1\right)
\label{omega2brane}\\
\Omega &\equiv& H_1 \Mt^2 \left[H_1 + \frac{2}{\Mt} \left(\beta -
\frac{\alpha}{6} H_1^2 \right) + \frac{\alpha}{3\Mt^2}H_1 \left(2\beta
- \mt^2 - \frac{7\alpha}{12} H_1^2 \right) \right].
\label{omega2shape}
\end{eqnarray}

The function $\omega^2$ is well approximated by $\omega_{\rm b}^2$
only near the brane and the expansion (\ref{omega2near}) is no longer
valid at large distance where the exponential term dominates [see
Eq.~(\ref{omega2})]. Once the fixed asymptotic values of the
Higgs and warp factor are reached, the frequency (\ref{omega2})
behaves as
\begin{equation}\label{omega2infty}
\omega^2\sim\omega^2_\infty(\varrho)\qquad\hbox{with}
\qquad\omega^2_\infty = \mt^2 \de^{2|\varrho|/\sqrt{6}} - \left(\Mt
+ \frac{1}{2\sqrt{6}} \right)^2.
\end{equation}
The analytical estimate of the function $\omega^2$ is thus obtained by
matching the two limiting asymptotic behaviors (\ref{omega2near}) and
(\ref{omega2infty}), respectively $\omega_{\rm b}^2$ closes to the
brane and $\omega^2_\infty$ far from it, as
\begin{equation}\label{omega2estim}
\omega^2(\rho)=\left\lbrace
\begin{array}{cc}
\omega_{\rm b}^2 & |\rho|<\rho_{_{\rm m}}\\
\omega_{\infty}^2 & |\rho|>\rho_{_{\rm m}}
\end{array}\right.
\end{equation}
The dimensionless matching distance $\varrho_{_{\rm m}}$ has to be
solution of
\begin{equation}
\omega^2_{_{\rm b}}(\varrho_{_{\rm m}})=\omega^2_\infty(\varrho_{_{\rm m}}),
\end{equation}
in order to get a continuous function. Note that, because of the
symmetry on both sides of the wall, we can assume $\varrho_{_{\rm
m}}>0$ without lack of generality. For large values of $\Mt$, and
using Eq.~(\ref{omega2shape}), we get
\begin{equation}
\label{shapeapp}
\sqrt{\Omega} \sim \Mt H_1,
\end{equation}
leading to
\begin{equation}\label{rhomatch}
\varrho_{_{\rm m}} \sim \frac{1}{H_1}.
\end{equation}
As it turns out, the faster the asymptotic solution is reached, the
better the approximation works. This is the case in particular for
$H_1>1$,

The exact (numerically integrated) effective potential and its
approximation are compared in Fig.~\ref{figapprox}. The global shapes
are effectively the same, and in spite of uncertainties at
intermediate regions due to this crude approximation, it is reasonable
to expect the same fermion physical behaviors in both potentials. Note
that for (cosmologically favored) higher value of $\alpha \beta$, the
Higgs field and the warp factor reach more rapidly their asymptotic
values leading thus to a better agreement between the two potentials,
as can be seen on Fig.~\ref{figapprox}.

\subsection{Determination of the bound states}\label{subsec_bound}

Given the approximate frequency (\ref{omega2estim}), the equation of
motion for the left-handed bulk spinor component $\Uh_{\rm L}$ reduces
to
\begin{eqnarray}
\varrho>\varrho_{_{\rm m}}: & \qquad& \left[\partial_\varrho^2 +
\mt^2 \de^{2 |\varrho|/\sqrt{6}} - \left(\Mt + \frac{1}{2\sqrt{6}}
\right)^2\right]\Uh_{\rm L}=0,\label{approxfar}\\
\varrho<\varrho_{_{\rm m}}: & \qquad& \left[\partial_\varrho^2+
\omega_0^2-\Omega \varrho^2\right]\Uh_{\rm L}=0,\label{approxnear}
\end{eqnarray}
with the requirement that $\Uh_{\rm L}$ and its derivative are
continuous at $\varrho_{_{\rm m}}$. Note also that we consider only
the case $\varrho>0$, physics on both sides of the brane being
completely symmetric under the transformation $\varrho\to
-\varrho$. Let us consider the solutions in each region separately.
\begin{figure}[t]
\begin{center}
\epsfig{file=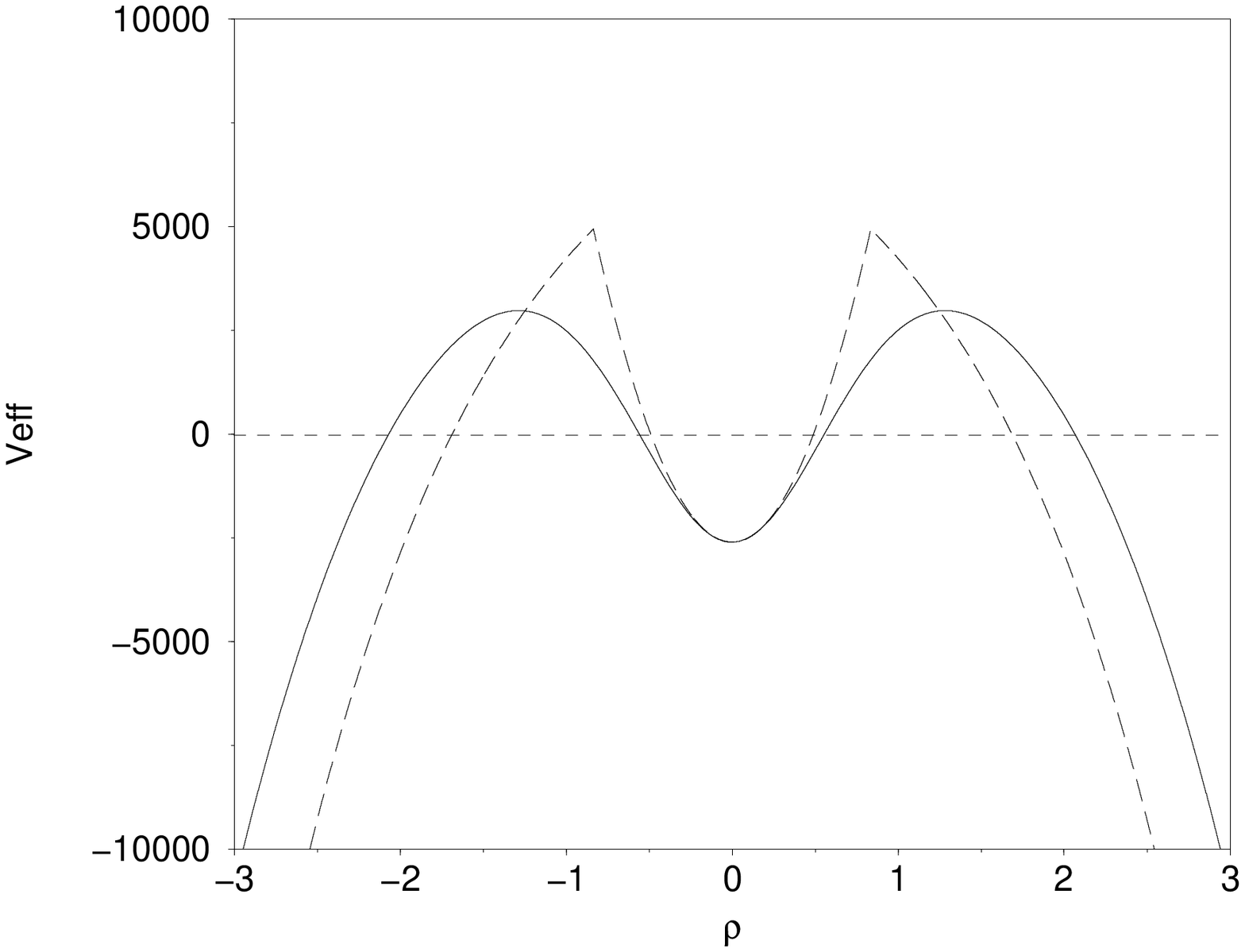,width=8cm}
\epsfig{file=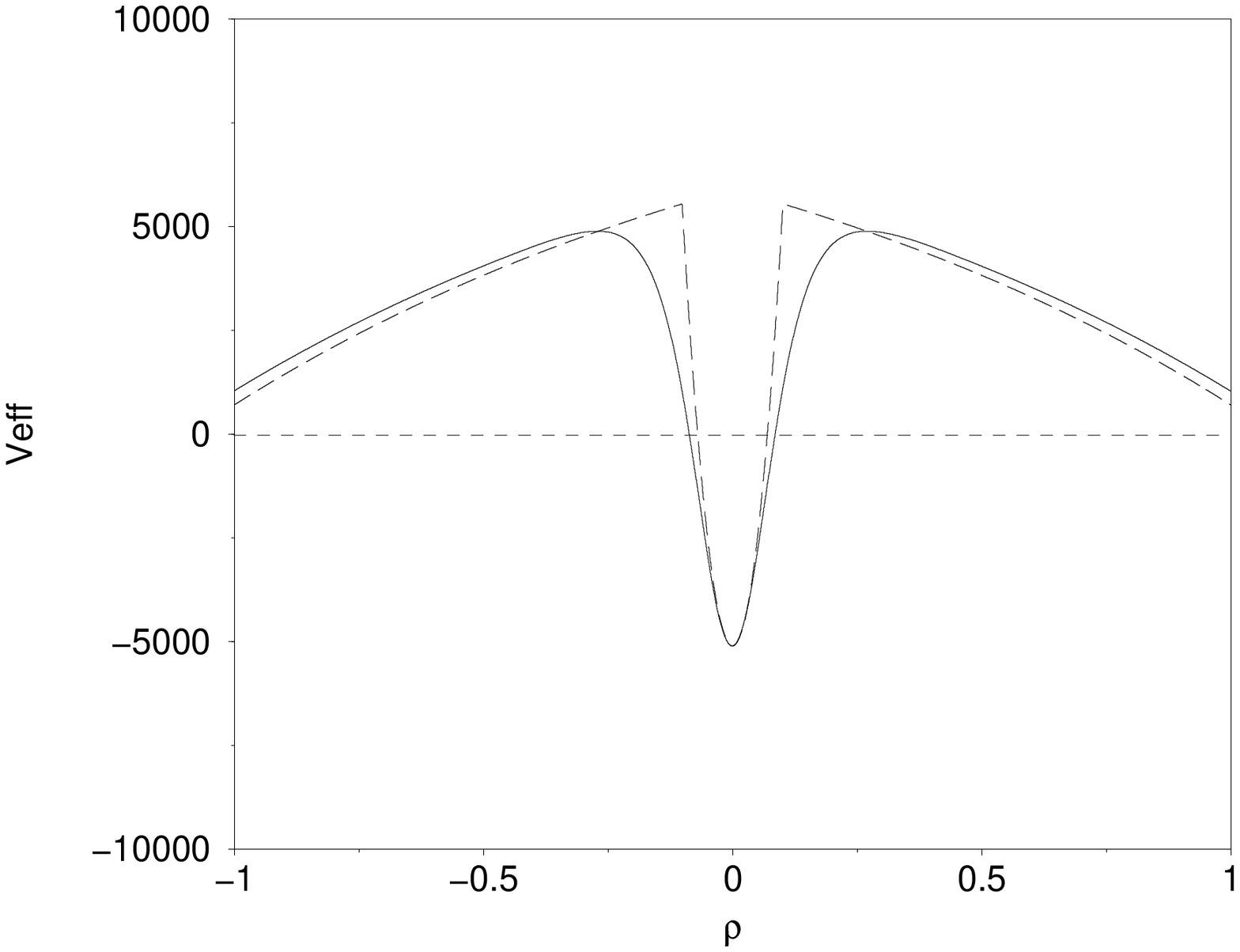,width=8cm}
\caption{The effective potential $V_{_{\rm eff}}=-\omega^2$ and the
approximate analytic effective potential (dashed curve) obtained from
the matching of its two asymptotic expansions near the brane and at
infinity. On the left panel, the parameters are the same as on
Fig.~\ref{figveff}, i.e., $\Mt= 100$ and $\mt = 0.5 \Mt$. As the
asymptotic solution is not reached very close to the brane, the
approximation is rather poor and the two solutions are not in good
agreement at intermediate regions. On the right panel, we use the new
parameter values $\Mt= 100$ and $\mt = 0.6 \Mt$, obtained for $\alpha
= 0.1865$ and $\beta = 53.62$, a much better approximation is obtained
around the potential barrier provided the Higgs field and warp factor
reach their vacuum value rapidely, i.e., for larger value of $\alpha
\beta$. As on Fig.~\ref{figveff}, the parameters are chosen to
illustrate the point and do not necessarily correspond to existing
bound states.}
\label{figapprox}
\end{center}
\end{figure}

\begin{itemize}
\item $\varrho>\varrho_{_{\rm m}}$: we introduce the new variable
\begin{equation}\label{defz}
z\equiv\sqrt{6}\mt\de^{|\varrho|/\sqrt{6}},
\end{equation}
in terms of which Eq.~(\ref{approxfar}) reduces to a standard Bessel
differential equation
\begin{equation}\label{bessel}
\left[\frac{\dd^2}{\dd z^2} + \frac{1}{z}\frac{\dd}{\dd z} + \left(1 -
\frac{\mut^2}{z^2}\right)\right]\, \Uh_{\rm L} = 0,
\end{equation}
the order of which being given by
\begin{equation}
\label{defmut}
\mut=\sqrt{6}\Mt+\frac{1}{2}.
\end{equation}
Since $\omega^2$ is positive at infinity, the asymptotic form of the
solution is necessary radiative, as was already pointed out in the
previous section.  The most general solution of the Bessel equation
(\ref{bessel}) is a linear superposition of Hankel functions. Since we
are interested only in ingoing waves in order to study a tunneling
process, the most general solution takes the form
\begin{equation}
\Uh_{\rm L}(z)=B \, H^{(1)}_{\mut}(z),
\end{equation}
where $H^{(1)}_{\mut}(z)$ is the Hankel function of the first kind,
propagating towards the brane at infinity~\cite{abramo}
\begin{equation}
H^{(1)}_{\mut}(z\rightarrow\infty) \sim \sqrt{\frac{2}{\pi z}}
\,\de^{i(z-\mut\pi/2-\pi/4)}.
\end{equation}
and $B$ is an arbitrary complex constant.
\item $\varrho<\varrho_{_{\rm m}}$: performing operations similar to
those of the previous case, we cast Eq.~(\ref{approxnear}) on the form
\begin{equation}\label{pcf}
\left[\frac{\dd^2}{\dd x^2} - \left(\frac{1}{4} x^2 + a \right)
\right] \Uh_{\rm L}=0,
\end{equation}
in which we have introduced the new variable and parameter
\begin{equation} x\equiv \left(4 \Omega\right)^{\frac{1}{4}} \varrho,\
\ \ \ \ a \equiv - \frac{\omega_0^2}{2 \sqrt{\Omega}}.
\label{defa}\end{equation}
The general solutions of Eq.~(\ref{pcf}) are the parabolic cylinder
functions, namely $U(a,x)$ and $V(a,x)$, of which $\Uh_{\rm L}(x)$ can
be expressed as linear superpositions. In the limit $x \gg |a|$, these
solutions scale as~\cite{miller}
\begin{eqnarray}\label{asymppcf}
U(a,x) \sim \de^{-\frac{1}{4}x^2} x^{-a-\frac{1}{2}}, & \quad
\textrm{and} \quad & V(a,x) \sim \sqrt{\frac{2}{\pi}}
\,\de^{\frac{1}{4}x^2} x^{a-\frac{1}{2}}.
\end{eqnarray}
Since we are interested in confined fermion states on the brane, only
the exponentially decreasing function is relevant, so that the general
solution near the brane reads
\begin{equation}
\Uh_{\rm L}(x) = A \, U(a,x),
\end{equation}
where $A$ is a complex integration constant.
\end{itemize}

The general solution $\Uh_{\rm L}(\varrho)$ for all $\varrho$ is
obtained by matching the two different solutions at
$\varrho=\varrho_{_{\rm m}}$. Since $\varrho_{_{\rm m}}$ corresponds
to the maximum positive value of the effective potential [see
Fig.~\ref{figapprox}], it is reasonable to consider that the Hankel
function at that point can be expanded around small values of their
argument with respect to their order~\cite{abramo}, i.e.,
\begin{equation}
\label{hankelmatch}
H^{(1)}_{\mut}(z_{_{\rm m}}) \sim -\frac{2^{\mut}}{\pi} \Gamma(\mut)
z_{_{\rm m}}^{-\mut},
\end{equation}
while the parabolic cylinder functions can be taken in their large
argument asymptotic limit (\ref{asymppcf}). This is the same kind of
approximation as that made to derive the effective
potential. Physically, the initial conditions on the brane, i.e.,
$\Uh_{\rm L}(0)$ and $\partial_\varrho\Uh_{\rm L}|_0$, are chosen in
such a way that the asymptotic exponentially growing function $V(a,x)$
contribution is everywhere negligible. Once these initial conditions
are fixed, they fully determine the solution on the other side of the
brane, i.e., for $x<0$. The asymptotic expansion (\ref{asymppcf}) can
be analytically extended to $-|x|=|x|\de^{i\pi}$ and
yields~\cite{gradsh}
\begin{equation}\label{pcfneg}
U(a,-|x|) \sim \de^{-\frac{1}{4}|x|^2} |x|^{-a-\frac{1}{2}}
\de^{-i\pi\left(a+\frac{1}{2}\right)}.
\end{equation}
Thus, once $\Uh_{\rm L}(0)$ and $\partial_\varrho\Uh_{\rm L}|_0$ are
fixed, the matchings between $H^{(1)}_{\mut}(z_{_{\rm m}})$ and
$U(a,-|x_{_{\rm m}}|)$ on one side, and $U(a,|x_{_{\rm m}}|)$ on the
other side fully determine the bulk component $\Uh_{\rm L}$ for all
$\varrho$.

The last constraint comes from Eq.~(\ref{orderzerour}) determining the
right-handed spinor bulk function. It is well defined if and only if
both $\Uh_{\rm L}$ and $\partial_\varrho \Uh_{\rm L}$ are not
singular. In fact, the derivative of the parabolic cylinder function
$U(a,x)$ is generally discontinuous at $x=0$. With the help of the
Wronskian of $U(a,x)$ and $U(a,-x)$~\cite{miller}
\begin{equation}
U(a,x) \frac{\dd U(a,-x)}{\dd x}- U(a,-x) \frac{\dd
U(a,x)}{\dd x}=\frac{2\pi}{\Gamma\left(\frac{1}{2}+a\right)},
\end{equation}
we can construct the derivative discontinuity at $x=0$. This is
\begin{equation}\label{jump}
\frac{\dd U(a,0^-)}{\dd x} - \frac{\dd U(a,0^+)}{\dd x}=2^{\frac{a}{2}
+ \frac{3}{4}} \frac{\Gamma\left(\frac{3}{4} +
\frac{a}{2}\right)}{\Gamma\left(\frac{1}{2} + a\right)},
\end{equation}
where we have used the particular value~\cite{miller}
\begin{equation}
\label{uzero}
U(a,0)=\frac{\sqrt{\pi}}{2^{\frac{a}{2} + \frac{1}{4}}
\Gamma \left(\frac{3}{4} + \frac{a}{2}\right)}.
\end{equation}
Imposing that the derivative of $\Uh_{\rm L}$ is continuous at $x=0$
results in imposing that the jump (\ref{jump}) vanishes.  This is the
case if and only if $a$ is solution of
\begin{equation}\label{nulwronsk}
\frac{\Gamma \left(\frac{3}{4} +
\frac{a}{2}\right)}{\Gamma\left(\frac{1}{2} + a\right)} = 0.
\end{equation}
Since $\Gamma$ is singular for negative integer arguments, this
condition is satisfied only for
\begin{equation}
\label{quantization}
-a-\frac{1}{2}=2n,
\end{equation}
where $n$ is a positive integer. Note that $-a-1/2$ cannot be odd
since then the numerator of the Wronskian~(\ref{nulwronsk}) will also
be singular resulting in a finite derivative jump at $x=0$. The
condition (\ref{quantization}) shows that the trapped fermions on the
brane have necessarily discrete masses $\mt_n$ which read, using the
values of the parameters~(\ref{omega2brane}), (\ref{omega2shape}) and
(\ref{defa}),
\begin{eqnarray}
\mt^2_n & = & \Mt H_1 \left[(4n+1) \sqrt{1 + \frac{6\beta}{3\Mt H_1} +
\frac{24 \alpha \beta  + \alpha^2 H_1^2
\left[(4n+1)^2-5\right]}{36\Mt^2}} \right.
- \left. \left(1+\frac{\alpha H_1}{6\Mt}
\left[(4n + 1)^2 + 1 \right] \right) \right].
\end{eqnarray}
This mass spectrum is valid for $n>0$ since our derivation assumed
that $\mt>0$. In the limiting case where $\Mt\gg 1$, it reduces to the
much simpler form for the lowest masses
\begin{equation}\label{appspectrum}
\mt_n \sim  2 \sqrt{n} \sqrt{\Mt H_1}.
\end{equation}
On the other hand, $\mt^2$ cannot reach very large values since it is
necessary to have a potential barrier in order to have bound
states. From the expression of the effective potential
(\ref{omega2infty}), the barrier is found to disappear when
\begin{equation}
\omega_\infty^2(\varrho_{_{\rm m}},\mt_{\max}) \sim 0.
\end{equation}
Again in the limit where $\Mt\gg 1$, using the value (\ref{rhomatch})
of $\varrho_{_{\rm m}}$, one gets the maximum accessible reduced mass
$\mt_{\max}$ for $\mt$ as
\begin{equation}\label{defmtmax}
\mt_{\max} \sim \Mt \de^{-1/\sqrt{6} H_1}.
\end{equation}
The maximum number of distinct massive states trapped on the brane
can thus be estimated to be
\begin{equation}
\label{defnmax}
n_{\max} \sim {\mathrm{Int}}\left[\Mt \frac{1}{4 H_1}\de^{-2
/\sqrt{6}H_1} \right].
\end{equation}
For the parameters chosen in Fig.~\ref{figveff}, one obtains
$\mt_{\max} \sim 0.68 \Mt$ and there are $n_{\max}=11$ massive modes
trapped on the brane.

\subsection{Fermion tunneling rate}\label{subsec_tunel}

Since the effective potential becomes negative at infinity, the
massive modes trapped on the brane are subject only to a finite
potential barrier. They are in a metastable state and can tunnel from
the brane to the bulk. In this section we use our previous analytic
solution to estimate the tunneling rate. Would this rate be too high,
one would observe an effective violation of energy-momentum
conservation on the brane, i.e., in four dimension, thereby
contradicting observation.

Our starting point is the analytic solution for the left-handed
bulk function that was derived in the previous section
\begin{eqnarray}
\varrho<\varrho_{_{\rm m}}: &\qquad& \Uh_{\rm L}(\varrho)=A\,
U\left(a,\sqrt{2}\Omega^{1/4} \varrho\right), \\
\varrho>\varrho_{_{\rm m}}: &\qquad& \Uh_{\rm L}(\varrho)=B \,
H^{(1)}_{\mut}\left(\sqrt{6}\mt\de^{|\varrho|/\sqrt{6}}\right).
\end{eqnarray}
The transmission factor can easily be derived from the matching
conditions of the left-handed bulk fermion component at
$\varrho=\varrho_{_{\rm m}}$. First of all, $\Uh_{\rm L}$ has to be
continuous. Using the expansions~(\ref{asymppcf}) and
(\ref{hankelmatch}) and the value (\ref{quantization}) of the
parameter $a$, permits to find the relation
\begin{equation}\label{amppos}
A \, \de^{-\frac{1}{2}\sqrt{\Omega}\varrho_{_{\rm m}}^2}
\left(2\sqrt{\Omega}\right)^n \varrho_{_{\rm m}}^{2n} =
-\frac{i}{\pi}B \, \Gamma(\mut)
\left(\sqrt{\frac{3}{2}}\mt_n\right)^{-\mut} \de^{-\mut
\varrho_{_{\rm m}}/\sqrt{6}}
\end{equation}
between the coefficients $A$ and $B$. Making use of the
expression~(\ref{rhomatch}) for $\varrho_{_{\rm m}}$ yields
\begin{equation}\label{baba}
\frac{B}{A}= \frac{i \pi}{\Gamma(\mut)}
\left(\sqrt{\frac{3}{2}}\mt_n\right)^{\mut}
\left(\frac{\mut^2}{3\sqrt{\Omega}}\right)^n
\exp{\left[\frac{\mut^2}{12\sqrt{\Omega}}\right]}.
\end{equation}

Assuming, as above, that $\Mt \gg 1$ (so that $\mut\gg1$) we can
expand $\Gamma(\mut)$ as
\begin{equation}
\Gamma(\mut) \sim \mut^{\mut-1/2} \de^{-\mut} \sqrt{2 \pi},
\end{equation}
so that, using expressions~(\ref{shapeapp}), (\ref{rhomatch}) and
(\ref{defmut}) respectively of $\Omega$, $\varrho_{_{\rm m}}$, and
$\mut$, we get that $B/A$ is approximated by
\begin{equation}\label{rappamp}
\frac{B}{A} \sim i \sqrt{\frac{\pi}{2}} \left(\frac{2}{H_1}\right)^n
6^{1/4} \Mt^{n+1/2} \exp{\left[- \sqrt{6}\Mt
\left(\ln{\frac{2\mt_{\max}}{\mt}} - 1 + \frac{1}{2 \sqrt{6}H_1}\right)
\right]}.
\end{equation}

The transmission coefficient from the brane to the bulk associated
with $\Uh_{\rm L}$ can thus be defined by
\begin{equation}\label{th}
\Th \equiv \frac{\Uh_{\rm L}(\mut)}{\Uh_{\rm L}(0)},
\end{equation}
with $\Uh_{\rm L}(\mut)=B H_\mut^{(1)}(\mut)$, is evaluated at the
turning point $\varrho=\mut$ where the spinor bulk component begins to
propagate freely. Using the behaviour (\ref{uzero}) of the function at
the origin, the ratio (\ref{rappamp}) and the properties~\cite{abramo}
of the Hankel function
\begin{equation}
H^{(1)}_{\mut}(\mut)=\left(\frac{4}{3}\right)^{2/3}
\frac{\de^{-i\pi/3}}{\Gamma(2/3)}
\, \mut^{-1/3},
\end{equation}
the transmission coefficient (\ref{th}) reduces to
\begin{eqnarray}\label{transmh}
\Th & \sim & \frac{2^{4/3}\Gamma(1/2-n)}{3^{7/12} \Gamma(2/3)} \de^{i\pi/6}
H_1^{-n} \Mt^{n+1/6}
\exp{\left[- \sqrt{6}\Mt \left(\ln{\frac{2\mt_{\max}}{\mt}} - 1 +
\frac{1}{2\sqrt{6} H_1}\right) \right]}.
\end{eqnarray}
It follows, using the definition~(\ref{defuh}), that the probability
for a trapped particle on the brane to tunnel to the bulk is given by
\begin{equation}
\label{probtunnel}
\Pb\equiv|\Tt|^2=\de^{2\mut/\sqrt{6}} \, |\Th|^2.
\end{equation}

The characteristic time for a fermionic mode trapped
on the brane can be roughly estimated by
\begin{equation}\label{timecar}
\tau_{_{\rm b}} \sim y_{_{\rm b}}=\frac{\varrho_{_{\rm
b}}}{\sqrt{|\Lambda|}},
\end{equation}
where $y_{_{\rm b}}$ represents the typical length, in the fifth
dimension, felt by a particle on the brane. As can be seen on
Fig.~\ref{figbulkradiate}, the spinor bulk components are
exponentially damped as soon as the effective potential becomes
positive. Thus $\varrho_{_{\rm b}}$ can be estimated by the solution
of $\omega_{_{\rm b}}^2(\varrho_{_{\rm b}})=0$ so that, keeping in
mind that $\mt<\Mt$,
\begin{equation}
\label{varrhob}
{\varrho_{_{\rm b}}}^{-1} \sim \sqrt{H_1 \Mt}.
\end{equation}
The life-time $\tau_n$ of a fermionic bound state on the brane
labelled by $n$
\begin{equation}
\tau_n=\frac{\tau_{_{\rm b}}}{\Pb}
\end{equation}
can be estimated by
\begin{eqnarray}\label{lifetime}
\tau_n & \sim & \frac{3^{7/6} \, \Gamma^2(2/3)}{2^{11/6} \,
\Gamma^2(1/2-n)}H_1^{2n -1/2} \frac{\Mt^{-2n - 5/6}}{\sqrt{|\Lambda|}}
\, \exp{\left(2\sqrt{6}\Mt\left[ \ln{\frac{2\mt_{\max}}{\mt_n}} - 1 -
{1\over \sqrt{6}} + \frac{1}{2\sqrt{6}H_1}\right]\right)}.
\end{eqnarray}

We recall that, due to the approximations performed in the previous
derivation, this estimate is valid only for
$\mt\ll\mt_{\max}$. Nevertheless, the argument in the exponential
amplifies
the transition from bound states to tunneling ones for masses $\mt
\sim \mt_{\max}$, as intuitively expected. An order of magnitude of
the minimal coupling constant $\Mt$ leading to stable bound states can
thus be estimated by requiring that the lowest massive state does not
tunnel,
\begin{equation}
\mt_1 <2 \mt_{\max} \exp{\left(-1 - {1\over \sqrt{6}} +
\frac{1}{2\sqrt{6}H_1}\right)}.
\end{equation}
Using the two values~(\ref{appspectrum}) and (\ref{defmtmax}),
this implies that
\begin{equation}
\Mt \gtrsim H_1 \exp{\left(2 + {2\over \sqrt{6}} +
\frac{1}{\sqrt{6}H_1} \right)}.
\label{frontier}\end{equation}
As a numerical application, for the Higgs and gravity parameters used
in Fig.~\ref{figveff}, we get $\Mt \gtrsim 25$.

\section{Numerical investigation}
\label{sec_num}

Numerically, it is simpler and more convenient to solve the first
order differential system (\ref{syst1}-\ref{syst2}). A Runge-Kutta
integration method was used, on both sides of the brane. In order to
suppress the exponential growth, we integrate from the turning point,
$\varrho=\mut$, where the solution begins to propagate freely, toward
the brane. In this way, we get only $U(a,x)$ near the brane. The
radiative solution for $\varrho>\mut$, is simply obtained by
integrating from the turning point toward infinity, with initial
conditions determined by the matching with the exponential decreasing
solution near the wall.  The same method is used on the other side of
the brane, but this time, by means of the last free parameter, we
impose the continuity of one bulk spinor component on the brane
($\Ut_{\rm L}$ say). Generally, the other bulk spinor component will
be discontinuous at $\varrho=0$, as expected from the analytical study
since $U'(a,0)$ is generally discontinuous. The mass spectrum is thus
obtained by requiring the continuity of $\Ut_{\rm R}$ on the brane.

\begin{figure}
\begin{center}
\epsfig{file=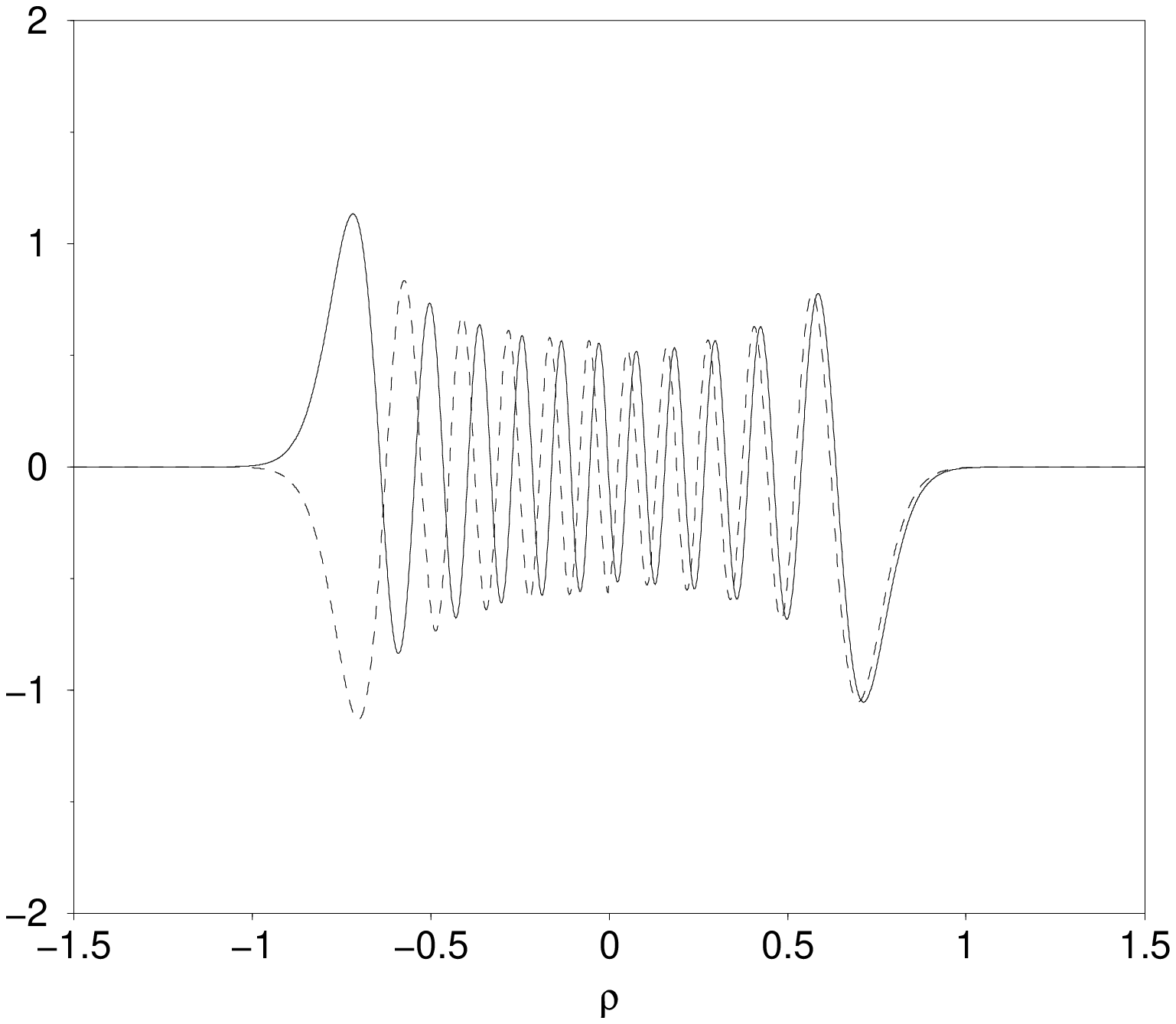,width=7cm}
\epsfig{file=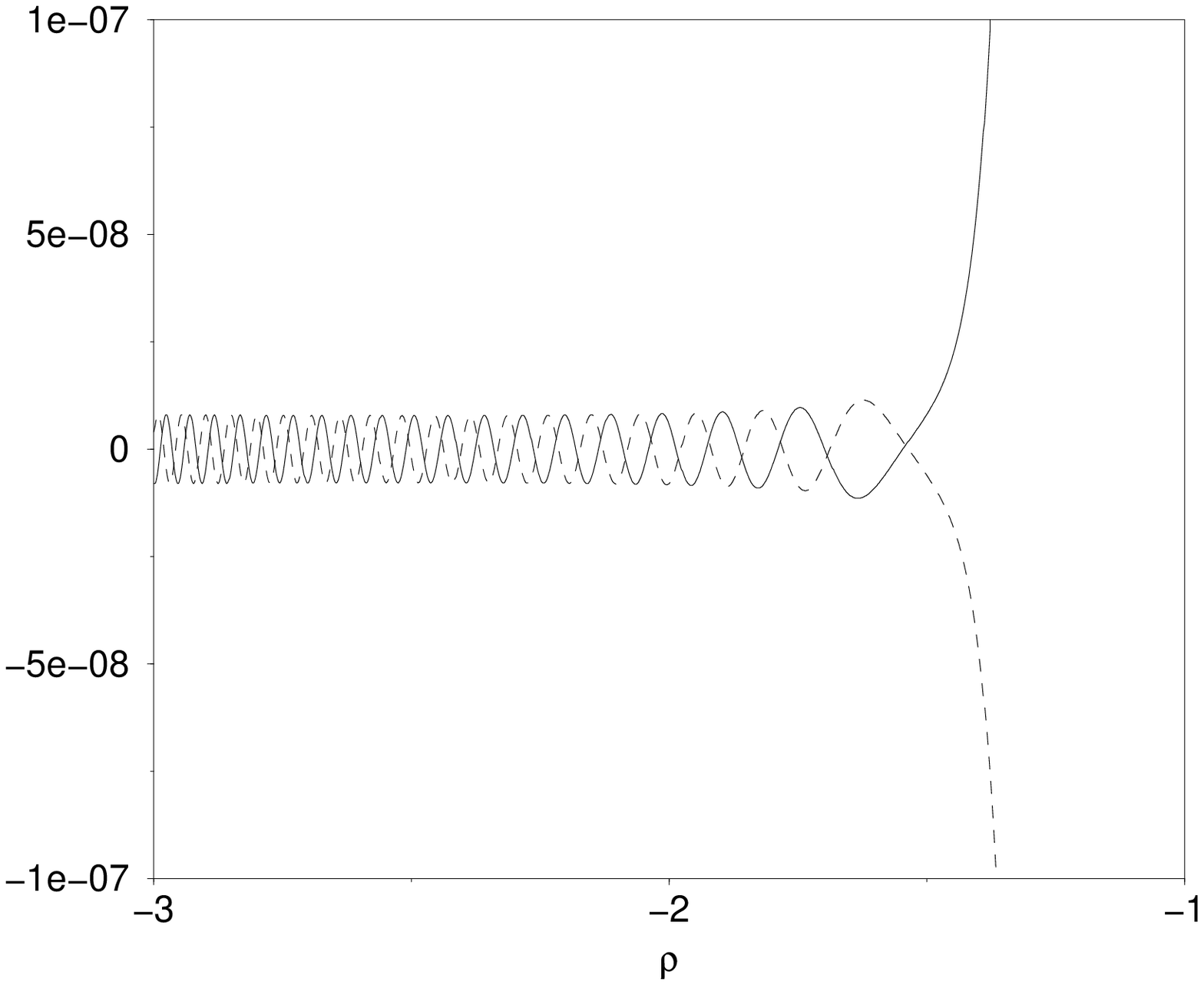,width=7.6cm}
\caption{[Left] The right (dashed line) and left (solid line) bulk
spinor components $\Ut_{\rm R}$ and $\Ut_{\rm L}$, as functions of the
dimensionless distance $\varrho$ to the brane, for the heaviest
massive state ($n=11$) and $\Mt=100$. [Right] Zoom near the turning
point and transition to the radiative behaviour which takes place at
a finite distance to the brane due to the tunneling of this mode from
the brane to the bulk.}
\label{figbulkradiate}
\end{center}
\end{figure}

\begin{figure}
\begin{center}
\epsfig{file=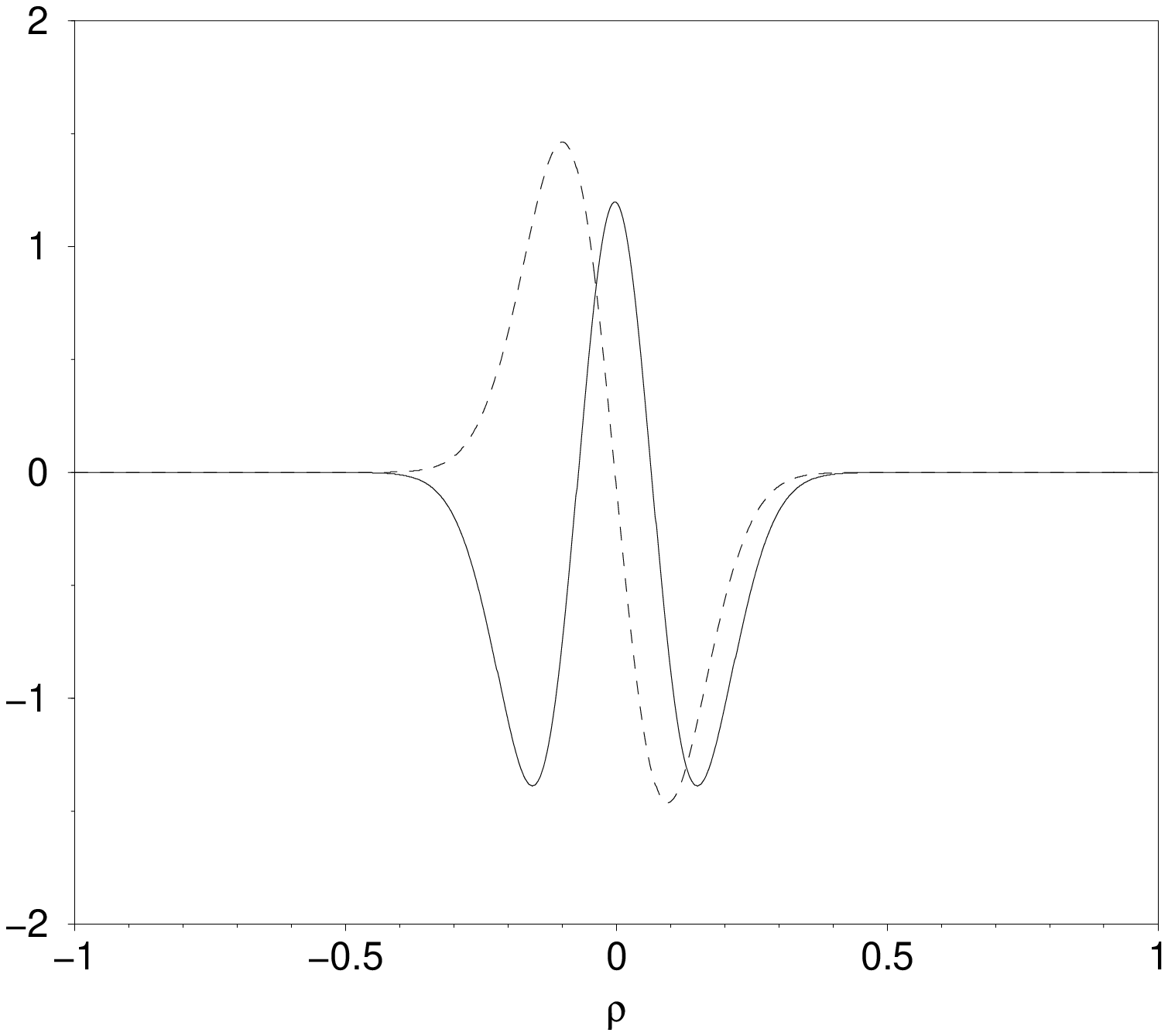,width=7cm}
\epsfig{file=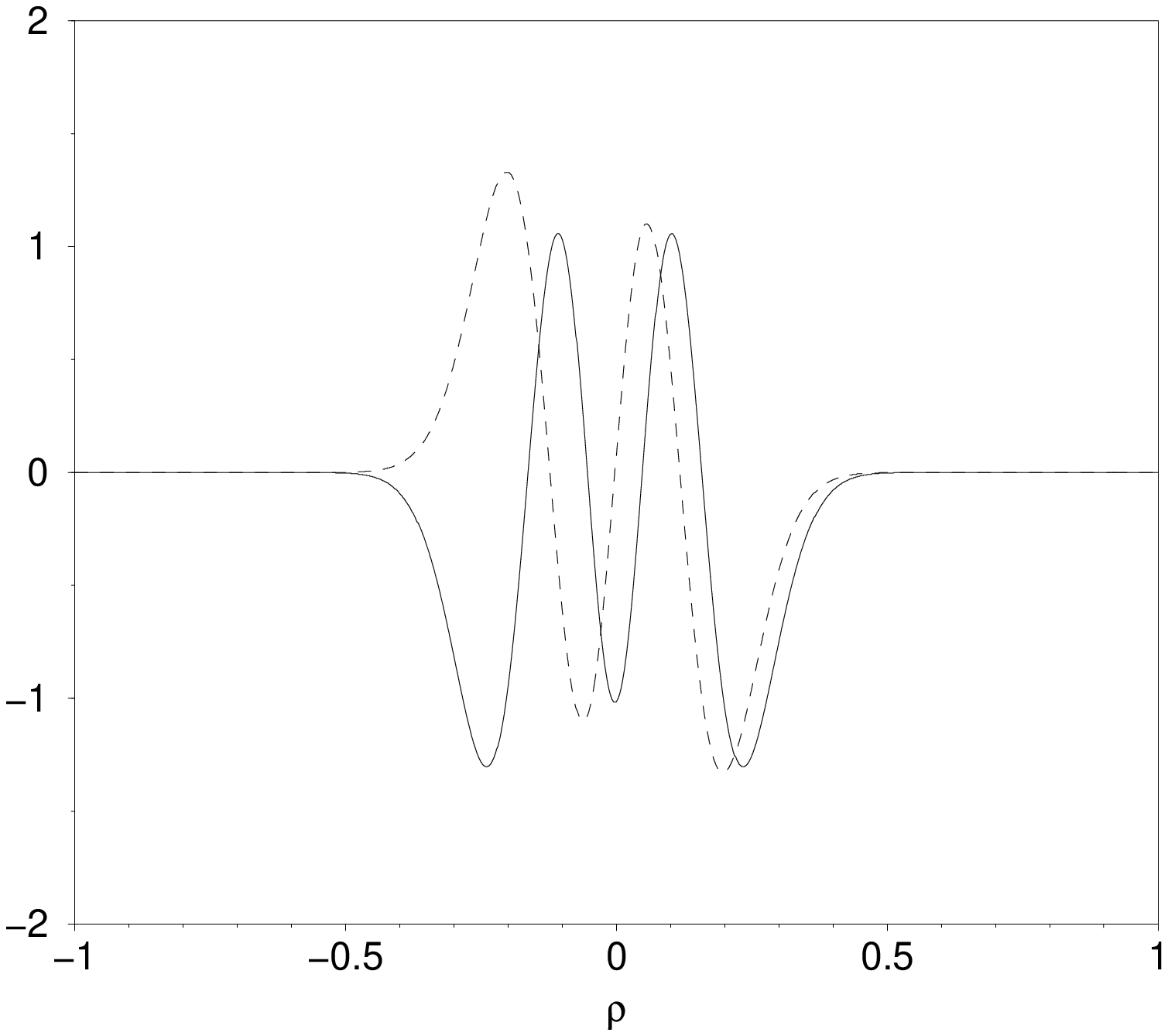,width=7cm}
\epsfig{file=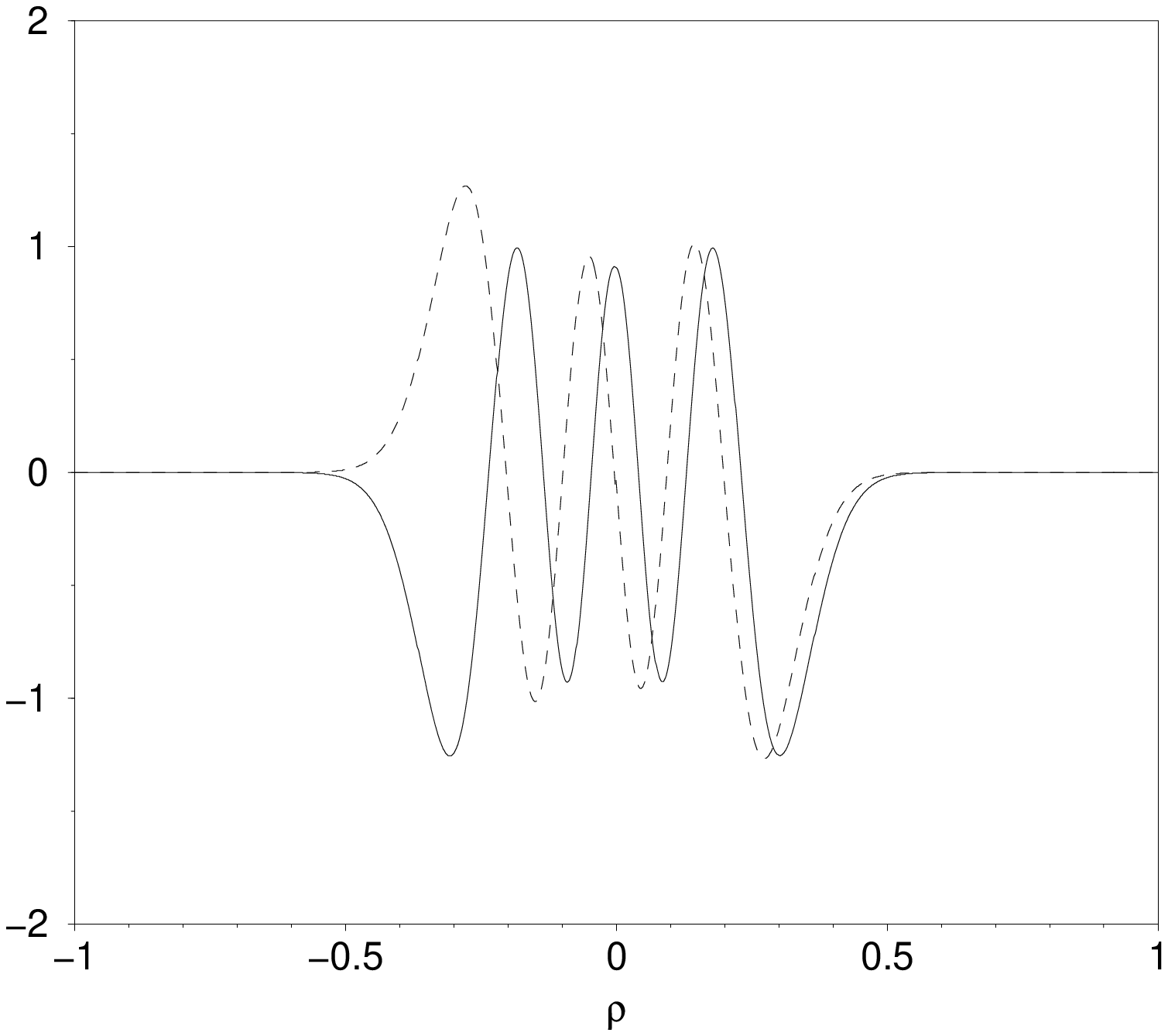,width=7cm}
\epsfig{file=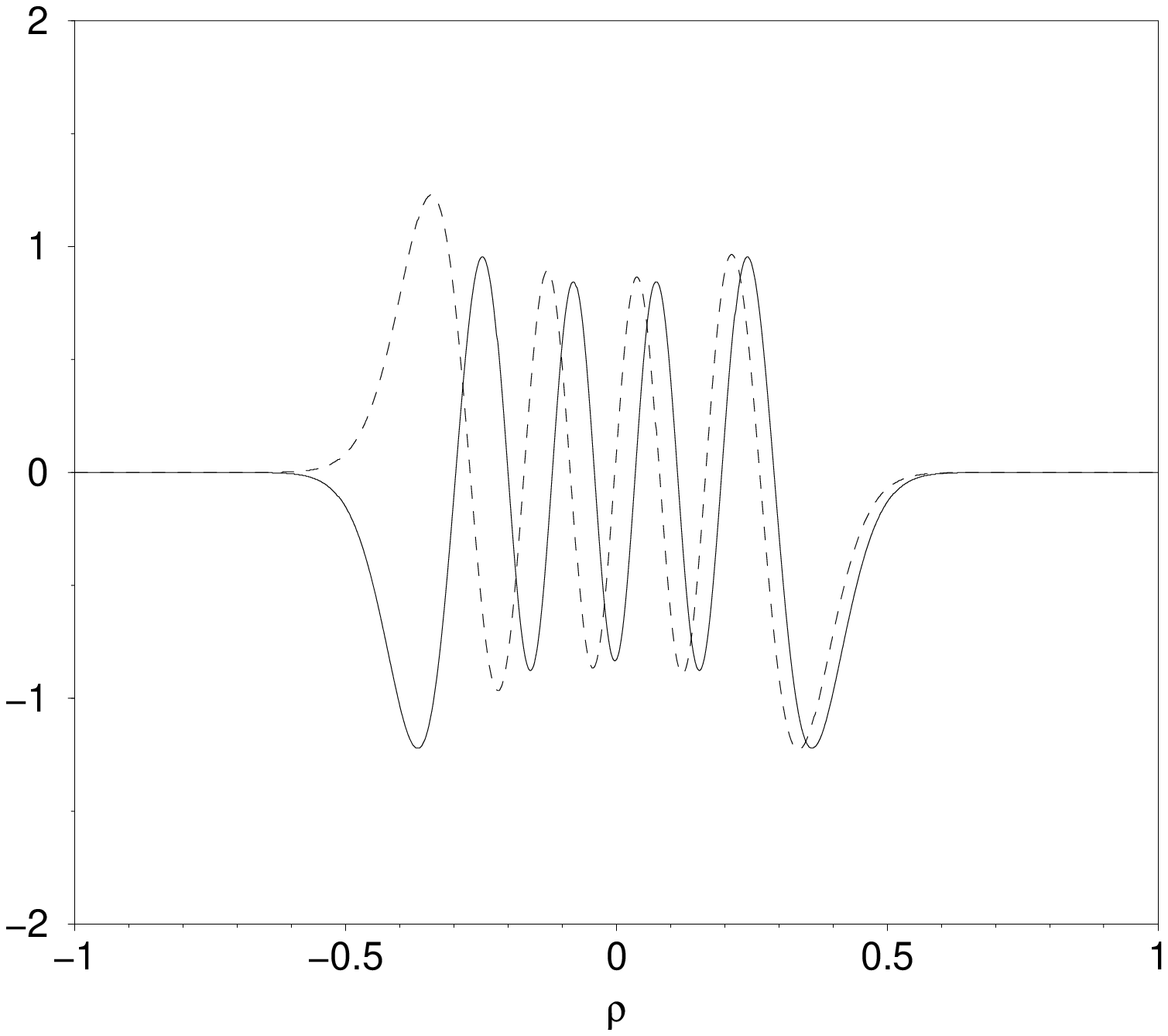,width=7cm}
\epsfig{file=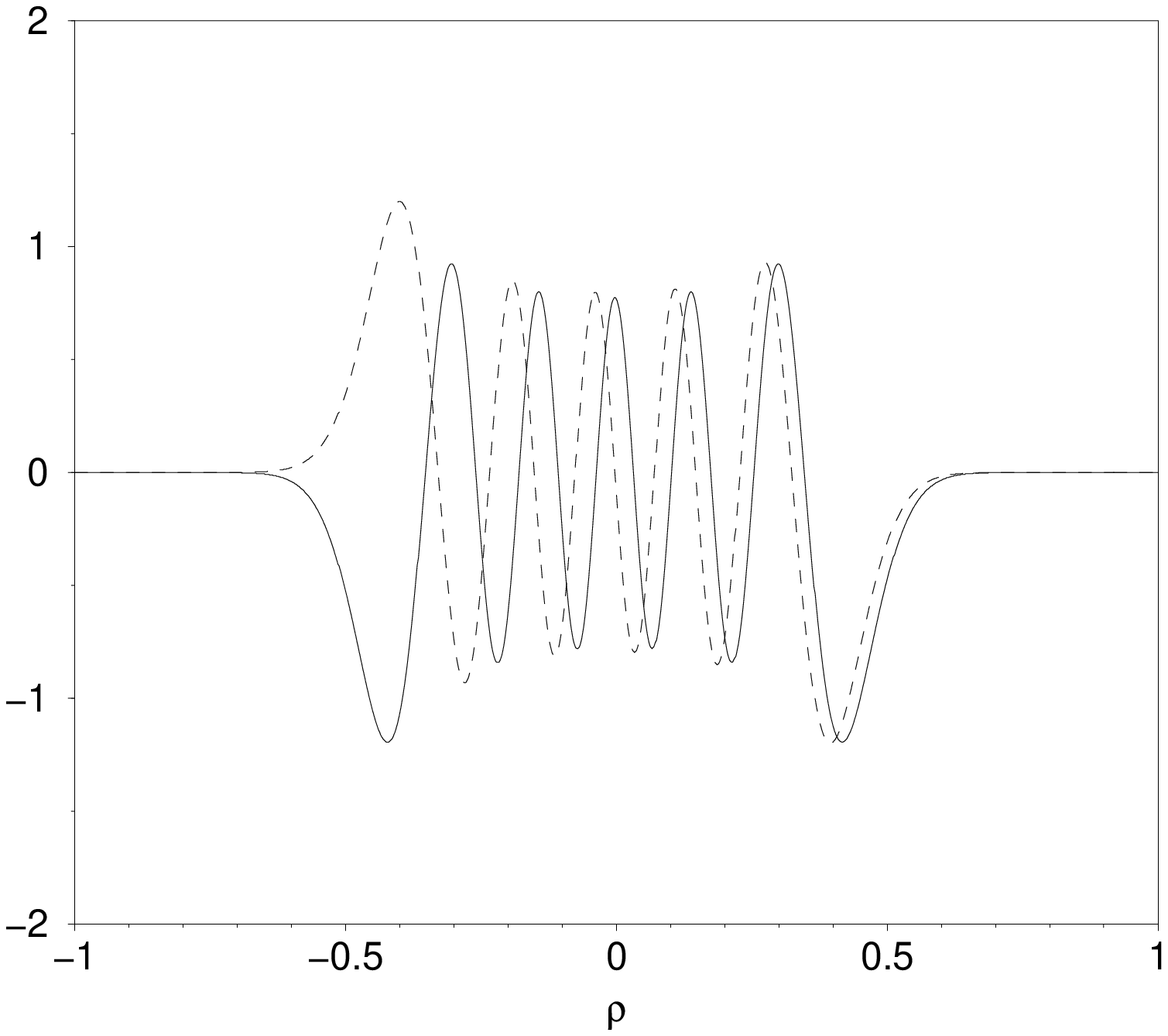,width=7cm}
\epsfig{file=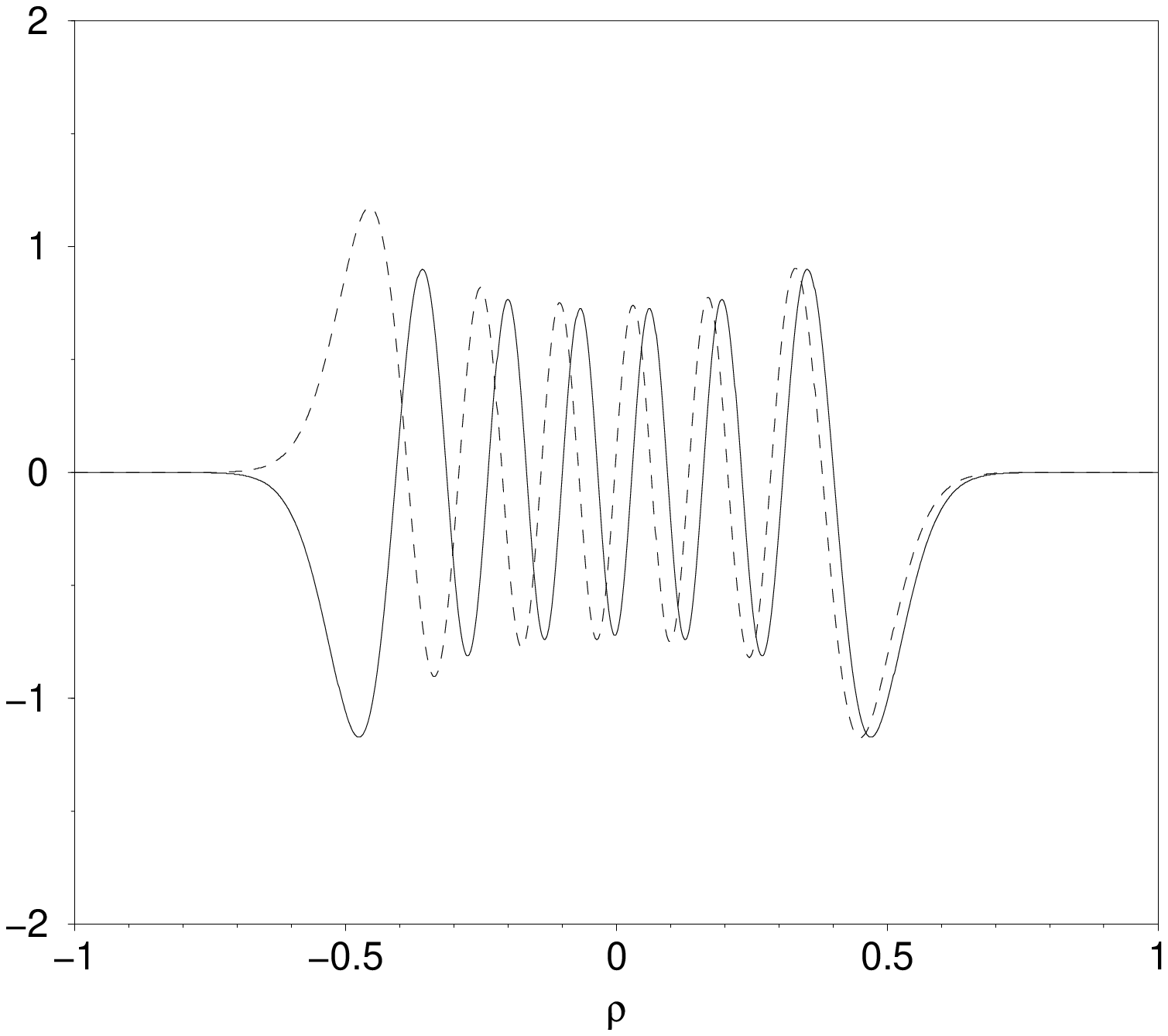,width=7cm}
\caption{The right (dashed curve) and left bulk spinor components,
$\Ut_{\rm R}$ and $\Ut_{\rm L}$, as functions of the dimensionless
distance $\varrho$ from the brane for the lightest massive bound
states. They have been computed for $\Mt=100$ and in the Higgs and
warp factor profiles obtained with parameter of
Fig.~\ref{figveff}. The numerical values of the corresponding reduced
masses are reported on table~\ref{table}.}
\label{figbulkcomp}
\end{center}
\end{figure}

The bulk spinor components computed this scheme have been plotted for
the first massive modes trapped on the brane in
Fig.~\ref{figbulkcomp}, for $\Mt=100$. The lowest mass is numerically
found to be $\mt_1\sim 0.209 \Mt$ and was estimated analytically,
from~(\ref{appspectrum}), to be $\mt_1 \sim 0.210 \Mt$ leading to a
precision of 0.5\% for the analytical estimate. The second mass is
numerically found to be $\mt_2 \sim 0.291\Mt$, which has to be
compared to its analytical estimate $\mt_2 \sim 0.295 \Mt$. Again, the
precision of this estimate is of about 1\%. As predicted
from~(\ref{defnmax}), there are $n_{\max}=11$ massive bound states the
lightest masses of which are summed up in table~\ref{table}. On
Fig.~\ref{figbulkradiate}, we plot the last $n=11$ trapped mode; it
has a tiny radiative component, as expected for a tunneling mode.

\begin{table}
\centering
\begin{tabular}{c|c|c|c|c|c|c|c c}
  $n$                 &   1   &   2   &   3   &  4    &   5   & 6 &
  \dots & 11\\
  $\mt_n$ (numerical) & 0.209 & 0.291 & 0.353 & 0.402 & 0.444 & 0.480
  & \dots & 0.593\\
  $\mt$ (estimates)   & 0.210 & 0.295 & 0.359 & 0.412 & 0.458 & 0.499
  & \dots & 0.657\\
  precision (\%)      &  0.5  &  1.3  &  1.7  & 2.4   &   3   &  3.8 &
  \dots & 9.7\\
\end{tabular}
\caption{Comparison of the numerical values and analytical estimates
of the first six bound states reduced mass $\mt$, together with the
heaviest mode, computed for $\Mt=100$.}\label{table}
\end{table}

In conclusion, the numerics confirm that the approximations of the
previous section and our estimates are accurate up to 1\%-10\% (see
table~\ref{table}).

\section{Discussion and conclusions}\label{sec_concl}

We shall now discuss the cosmological constraints existing on the kind
of model we have been considering here. Most of these constraints come
from brane models in which the wall structure is replaced by an
infinitely thin four dimensional layer. As discussed in the previous
sections, such an approximation is equivalent, within our framework,
to asking that the combination $\alpha\beta$ be much larger than
unity. In this limit, equivalent to the large $\beta$ limit since,
from Eq.~(\ref{relab}), $\alpha\beta\sim {4\over 3} \beta$, we can
replace the stress-energy tensor~(\ref{tmunu}) by the effective four
dimensional surface distribution
\begin{equation} T^{^{\rm eff}}_{\mu\nu} = T_\infty g_{\mu\nu} \delta
(y),\end{equation} whose isotropic tension $T_\infty$ is obtained by
integration in the transverse direction to yield
\begin{equation} T_\infty = \sqrt{|\Lambda|} \eta^2 \int \dd\varrho\,
\de^{-6\sigma(\varrho)} \left[ {1\over 2} \Phi'^2 + 2 V \right] \equiv
\sqrt{|\Lambda|} \eta^2 \xi (\beta) \end{equation}
where the function $\xi(\beta)$ can be expressed as
\begin{equation} \xi = \int \dd\varrho\,
\de^{-6\sigma(\varrho)} \left[ {6S^2-1\over \alpha} + 2 \beta (H^2-1)^2
\right]. \end{equation}

\begin{figure}
\begin{center}
\vskip3mm
\epsfig{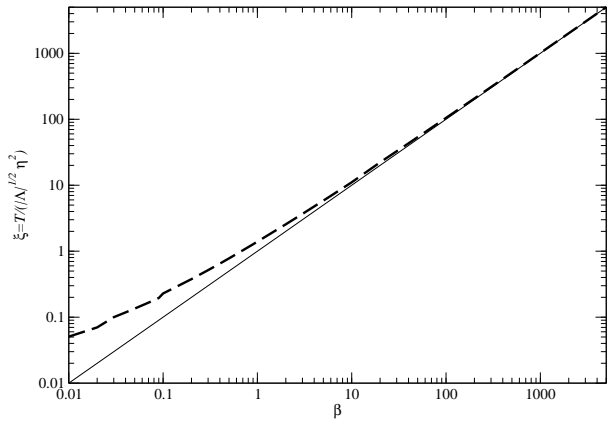}
\vskip6mm
\caption{Effective four dimensional brane tension, in units of
$\sqrt{|\Lambda|} \eta^2$, of the domain wall as a function of $\beta$
(dashed line). It is clear that in the thin brane limit,
$\xi\simeq\beta$ (thin line).}
\label{Tinf}
\end{center}
\end{figure}

Fig.~\ref{Tinf} shows the variation of $\xi$ as a function of
$\beta$. In the limit $\beta\gg 1$, it is clear that
$\xi\simeq\beta$, so that the brane tension behaves as
\begin{equation} T_\infty \simeq {\lambda\eta^4\over 8\sqrt{|\Lambda|}},
\label{tension} \end{equation}
which will be used to derive the relevant cosmological
constraints. Note also that the discrepancy between the above formula
and the actual value of $T_\infty$ becomes important (more than 100~\%
error say) for $\beta \alt 0.1$, which is already rather far from the
thin brane limit usually considered.

\subsection{Investigation of the parameter space}

The model described in this article depends of five parameters, four
describing the spacetime and scalar field dynamics
$(G_{_5},\Lambda,\eta,\lambda)$ and one concerning the fermions
$(g_{_{\rm F}}$). With the domain wall structure assumed, only four of
these parameters are independent [see Eq.~(\ref{rellambda})].  It is
convenient to replace this set of parameters by the three mass scales
\begin{equation}\label{mass_scale}
m_{_5}\equiv G_{_5}^{-1/3},\qquad
m_{_\Lambda}\equiv \frac{\sqrt{-\Lambda}}{6} \qquad
m_{_\infty}\equiv T_{_\infty}^{1/4},
\end{equation}
and the dimensionless parameter $\gamma_{_{\rm F}}$. These parameters
are subject to a number of constraints, namely
\begin{enumerate}

\item the four dimensional gravitational constant must agree with its
observed value $G_{_4}\equiv m_{_4}^{-2}$ with $m_{_4}\sim
10^{19}\,$GeV.  Using the expression of the four dimensional Planck
mass in terms of the five dimensional analog and of the brane tension
gives~\cite{cosmo5D,carter,shiro}
\begin{equation}\label{rel4}
m_{_5}^3 \sim m_{_4}m_{_\infty}^2.
\end{equation}

\item The brane cosmological constant~\cite{cosmo5D,carter,shiro}
\begin{equation}
2\Lambda_{_4}=\Lambda+6\pi^4G_{_5}^2T_{_\infty}^2,
\end{equation}
must also agree with the standard observational bound
$m_{\Lambda_{4}}<10^{-60}m_{_4}$. This implies
\begin{eqnarray}\label{rel2}
m_{_\Lambda}=\frac{\pi^2}{\sqrt{6}}\frac{m_{_\infty}^4}{m_{_5}^3} &
\Rightarrow & m_\infty^4 \sim m_\Lambda m_5^3.
\end{eqnarray}
Note that in the limit $\beta\gg 1$, this relation is equivalent to
Eq.~(\ref{tension}). This means that this condition is readily
satisfied in the thin brane limit. At this point, it is worth
emphasizing that this is precisely the limit in which the analytic
approximation for fermion masses are the most accurate.

\item There must not be any deviation of the law of gravity on the
brane with respect to the inverse square Newton law above 1
millimeter~\cite{limit}. This implies~\cite{chung}
\begin{equation}\label{rel5}
m_{_\Lambda}\gsim 10^{-3}\,\mathrm{eV}.
\end{equation}

\item Finally, we require the fermion stress-energy tensor to be
negligible with the brane stress-energy, so that we impose that the
mass of the heaviest fermion is smaller than the brane mass scale. By
means of Eq.~(\ref{defmtmax}), this condition reads
\begin{equation}
m_{\max} \sim m_\Lambda \Mt \de^{-1/(\sqrt{6}H_1)} < m_\infty,
\label{mmax}\end{equation}
where $H_1$ ends up being function of $\beta$ only by means of
Eqs~(\ref{relab}) and (\ref{def_H1}),
\begin{equation}
\label{h1beta}
H_1^2 = 2\beta - \frac{6 \beta}{\sqrt{16 \beta + 9}}.
\end{equation}
In the limit $\beta\gg 1$, Eqs.~(\ref{mmax}) and (\ref{h1beta})
combine to give the constraint on the coupling constant
\begin{equation} \Mt < {m_\infty\over m_\Lambda}
\de^{1/\sqrt{12\beta}} \simeq {m_\infty\over
m_\Lambda}.\label{gammaf}\end{equation}

\end{enumerate}

It follows from the relations (\ref{rel4}) and (\ref{rel2}) that
$m_\infty^2 \sim m_\Lambda m_4$ so that the three mass scales must
satisfy
\begin{equation}\label{numval}
m_{_\Lambda} \gsim 10^{-3}\,{\rm eV},\qquad
m_{_\infty} \gsim 1 \,{\rm TeV},\qquad
m_{_5} \gsim 10^6\,{\rm TeV},
\end{equation}
which, together with Eq.~(\ref{gammaf}) and (\ref{cond_bounded})
yields
\begin{equation} {1\over 2\sqrt{6}} < \Mt \alt
10^{15}\left( {m_\infty\over 1\,\hbox{TeV}}\right)
,\label{cstgammaf}\end{equation} where, as discussed below
Eq.~(\ref{defuh}), the lower bound is very conservative.

Now, let us examine the stability of the fermion confinement on the
brane and the restriction on their life-time imposed by the previous
conditions. We require, at least, one massive bound state to have a
life-time longer than the age of the Universe, i.e.,
\begin{equation}\label{t_u}
\tau_1 > \tau_{\mathrm{univ}} \sim 10^{17} \mathrm{s}.
\end{equation}
Using Eq.~(\ref{lifetime}), one roughly gets
\begin{equation}
m_\Lambda\tau_1 \sim \Upsilon(\beta,\Mt)\equiv H_1^{3/2} \Mt^{-17/6}
\exp{\left(2\sqrt{6}\Mt \left[\frac{1}{2} \ln{\frac{\Mt}{H_1}} - 1 -
1/\sqrt{6} - \frac{1}{2\sqrt{6}H_1}\right]\right)}.
\end{equation}
The condition (\ref{t_u}) together with the former constraint
(\ref{numval}) can be written as
\begin{equation}
\Upsilon(\beta,\Mt)\agt 10^{30}.
\end{equation} 

\begin{figure}
\begin{center}
\vskip5mm
\epsfig{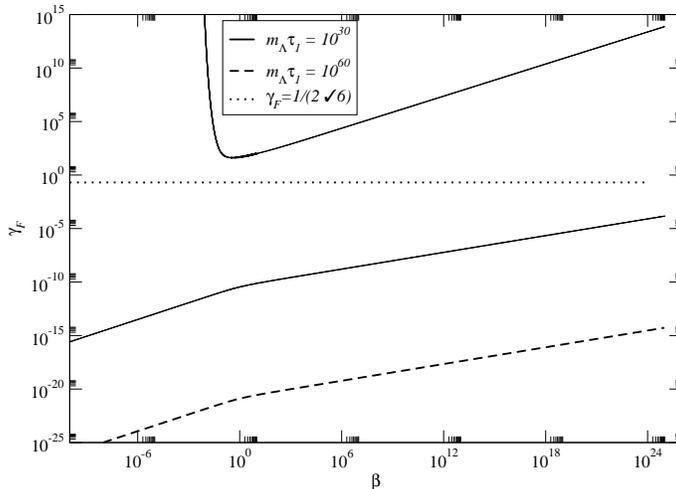}
\caption{Contour plot of the function $\Upsilon (\beta,\Mt)$ for
$\Upsilon =10^{30}$ [solid lines] and $10^{60}$ [dashed line], i.e.,
respectively a particle life-time of the order of the age of the
Universe, and of the proton life-time lower limit. Note that both top
curves are indistinguishable due to the exponential behavior of
$\Upsilon$ in $\Mt$. They turn out to be equivalent to the analytical
requirement given by Eq.~(\ref{frontier}). Points above the highest
curve and below the lowest curves satisfy the contraint. We also
superimpose the conservative constraint (\ref{cond_bounded}) necessary
for the existence of a massive bound state, the allowed region being
above the dotted line.}
\label{figlifetime}
\end{center}
\end{figure}

On Fig.~\ref{figlifetime}, we present the contour plot of the
dimensionless function $\Upsilon(\beta,\Mt)$, for $\Upsilon =10^{30}$
and $10^{60}$, which correspond respectively to a particle life-time
of the order of the age of the Universe, and of the proton life-time
lower limit. For $\beta \agt 1$, there are in principle two allowed
regions, corresponding to strong and weak coupling limits,
i.e. $\Mt\gg 1$ and $\Mt\ll 1$. However, the lower bound on $\Mt$,
which comes from the requirement that fermions are actually trapped on
the brane, pushes the weak coupling allowed region to very high values
of $\beta$, in practice $\beta \agt 7.5\times 10^{57}$ for $\Upsilon
=10^{30}$. Note also that this already rather extreme value is based
on the conservative estimate given by Eq.~(\ref{cond_bounded}).

For $\beta\alt 1$, the weak coupling region completely disappears,
while the strong coupling allowed region shrinks rapidely: for
$\beta\alt 8.5\times 10^{-3}$, the life-time cannot exceed the age of
the Universe because $\Mt\agt 10^{15}$. Considering $\beta\agt 1$
therefore turns out to be the relevant limit if one wishes to have
fermionic bound states living on the brane.

\subsection{Conclusion}

In this article, we have considered fermions coupled to a Higgs field
with a domain wall structure in a five dimensional anti-de~Sitter
spacetime. This domain wall can be thought of as a realization of a
brane universe.

After, studying the domain wall configuration, we solved the Dirac
equation and showed that there exists massive fermionic bound states
trapped on the wall. We develop both analytic approximation to compute
the mass spectrum and the tunneling time. This was compared to a full
numerical integration of the dynamical equations that revealed the
accuracy of our approximation scheme.

We recover the fact that massive fermions tunnel to the
bulk~\cite{dubovski}. Investigation in the parameter space shows that,
for models satisfying the cosmological constraints, the relevant
confinement life-time can be much greater that either the age of the
Universe or the proton life-time. This was made possible by the
derivation of the analytic estimate.

One of our central result is the derivation of an analytic mass
spectrum for fermions trapped on a brane-like four dimensional
spacetime. In particular, as could have been
anticipated~\cite{neronov}, it was shown that the allowed masses are
quantized, with a spectrum varrying, in the strong coupling limit, as
$\sqrt{n}$. Such a spectrum is indeed in contradiction with
experimental measurements of particle masses~\cite{pdg}, which is not
surprising given the simplicity of the model. It however opens the
possibility to build more realistic theories in which mass
quantization would stem naturally from extra dimensions.

\section*{Acknowledgments}
We thank J\'er\^ome Martin for numerous enlightning discussion, and in
particular for pointing to us Ref.~\cite{miller} on parabolic cylinder
functions. JPU thanks l'Institut d'Astrophysique de Paris for
hospitality while this work was carried out.

\end{document}